\documentclass[aps,preprint,showpacs,amsmath,amssymb]{revtex4}

\usepackage{epsf}
\usepackage{longtable}
\usepackage{amsfonts}
\usepackage{graphicx}
\usepackage{dcolumn}
\usepackage{bm}

\newcommand{\pa}{\partial}

\begin{document}

\title{
Third-and-a-half order post-Newtonian 
equations of motion for relativistic compact binaries using 
the strong field point particle limit}

\author{Yousuke Itoh}
\affiliation{Astronomical Institute, Tohoku University, \\
Sendai 980-8578, Japan}
\email{yousuke@astr.tohoku.ac.jp}

\date{\today}

\begin{abstract}
 We report our rederivation of the equations of motion 
for relativistic compact binaries through
 the third-and-a-half post-Newtonian (3.5 PN) order approximation to general
 relativity using the strong field point particle limit to describe self-gravitating
stars instead of the Dirac delta functional.  The computation is done
in harmonic coordinates.  
Our equations of motion describe the orbital motion of the binary
 consisting of spherically  symmetric non-rotating stars. 
The resulting equations of motion fully agree with the 3.5 PN 
equations of motion  derived in the previous works. 
We also show that the locally 
 defined energy of the star has a simple relation with its mass up to the 
 3.5 PN order. 
\end{abstract} 


\maketitle

\section{Introduction}

A relativistic compact binary (e.g., a neutron star binary) 
loses its orbital angular momentum by emitting gravitational waves 
and coalesces in the end. Such a system is a promising source 
for the gravitational wave detectors such as CLIO, GEO600, LCGT, LIGO,
TAMA300, and VIRGO \cite{Sathyaprakash:2009xs}. However, even with those 
advanced detectors,  a direct detection is not easy. Indeed, 
because the amplitude of gravitational wave from such a source is 
expected to be tiny at the Earth compared to the detectors' noises, an efficient
detection method is required. One method widely used is matched
filtering. When using this technique, it is known that the more accurately we know the 
shape of the signal, the larger the signal to noise ratio becomes. 
This in turn means that it is desirable to know the dynamics of the binary accurately when 
one hopes to increase the number of the detectable events.  This purpose can 
be achieved by using higher order post-Newtonian (PN) equations of motion for two 
point particles, because up to the last several orbits before the  
coalescence the binary components have moderately slow  
orbital velocities and are affected negligibly by tidal effects \cite{Bildsten:1992my,1997rggr.conf...33B}.  
In fact, it is suggested that at least a third order post-Newtonian 
correction in the equations of motion are necessary for extraction of
astronomical information of the sources \cite{Blanchet:2006zz}. To obtain higher order 
post-Newtonian corrections, mainly three methods are employed in the literature.

In the first method, the Arnowitt-Deser-Misner (ADM) Hamiltonian is
derived by means of a direct PN iteration of the Einstein equations 
\cite{Ohta:1973je,Ohta:1974pq,1985AnPhy.161...81S,1985GReGr..17..879D,1986GReGr..18..255S,Jaranowski:1996nv,JS98,Jaranowski:1998wy,Jaranowski:1999ye,Damour:2001bu,Konigsdorffer:2003ue}.   
Indeed, the ADM Hamiltonian in the ADM Transverse-Tarceless (TT) gauge
is completed up to the 3.5 PN order inclusively \cite{Jaranowski:1996nv,Konigsdorffer:2003ue}.  
In the second method, 
one assumes that the energy and the angular momentum fluxes at 
infinity are balanced by the corresponding loses of those in the binary
orbital motion. Here the known PN expressions for the energy and angular momentum 
fluxes at infinity are used. With this second method, the n PN (n = 2.5,
3.5, and 4.5) order corrections to the Newtonian equations of motion are
derived \cite{Iyer:1993xi,Iyer:1995rn,Blanchet:1996vx,Gopakumar:1997ng}. 
Finally in the third method, PN equations of motion are 
derived by a direct PN iteration of the Einstein equations \cite{Einstein:1938yz,Einstein:1940mt,1981PhLA...87...81D,Damour82,1983SvAL....9..230G,1985SvA....29..516K,1985PhRvD..31.1815T,Futamase87,BFP98,Blanchet:2000nv,IFA00,PW00,Blanchet:2000ub,IFA01,PW02,Itoh:2003fy,Itoh:2003fz,Itoh:2004dh,Nissanke:2004er,2007PhRvD..75f4017W,2007PhRvD..75l4025M}. 
(see also \cite{Damour1983,Schutz85,Damour1987,1997rggr.conf...33B,AsadaFutamase97,Blanchet:2006zz,Futamase:2007zz}
for reviews). With this method the equations of motion are completed up to the 3.5 PN order inclusively \cite{PW02,Nissanke:2004er}. 

This paper reports 
a rederivation of the 3.5 PN correction to the Newtonian acceleration
for a spherically symmetric non-rotating self-gravitating star using the same method as
that in our previous papers 
\cite{IFA00,IFA01,Itoh:2003fy,Itoh:2003fz,Itoh:2004dh} which derived the
corrections up to and including the 3 PN order (see \cite{Futamase:2007zz} for a review).  
We use the third method mentioned in the previous paragraph.  
However our method is different in the several aspects from those of the
previous works 
\cite{PW02,Nissanke:2004er} which also used the third method to derive
the 3.5 PN order corrections. 
Here we briefly mention the two most important points. 

The first point is regarding the way to describe the binary components as point particles. 
Among the previous works, Damour, Blanchet, Sch\"afer and their collaborators 
have used Dirac delta functionals to describe binary stars as   
point particles. However, divergent integrals appear   
when using Dirac delta functionals in general relativity.  
Then to regularize those divergences, for example, the work \cite{BFP98}
successfully used the Hadamard 
Partie Finie (HPF) regularization to derive the 2.5 PN equations of 
motion in harmonic coordinates. Blanchet and Faye have developed the
generalized HPF regularization \cite{BF00} in a Lorentz invariant
manner \cite{BF01} and derived the 3 PN 
correction in the same gauge except for one and only one numerical coefficient 
(denoted by $\lambda$) which could not be determined within their method
\cite{Blanchet:2000ub,Blanchet:2000nv}. Interestingly, two works \cite{deAndrade:2000gf,Damour:2000ni}    
have shown that this coefficient $\lambda$ actually corresponds 
to one of the undetermined 
coefficients ($\omega_{\rm static}$) reported previously by Jaranowski and Sch\"afer in their 
derivation of the 3 PN ADM Hamiltonian 
in the ADM Transverse-Traceless (TT) gauge \cite{JS98,Jaranowski:1998wy,Jaranowski:1999ye}. 
Indeed these latter works also used Dirac delta functionals and the HPF. 
It has become clear that the HPF is not an appropriate method for
regularization of the divergent integrals at the 3 PN order. 
Damour, Jaranowski and Sch\"afer \cite{Damour:2001bu} then used the dimensional
regularization to derive the 3 PN ADM Hamiltonian in the ADMTT gauge and
they finally completed the 3 PN correction that contains no undetermined
coefficient. The 3 PN equations of motion in harmonic coordinates were 
later derived using the combination of the HPF and the dimensional
regularization \cite{Blanchet:2003gy}. 
Their result physically agree with the result of \cite{Damour:2001bu}.  
Incidentally, there found no trouble when using the HPF 
at the 3.5 PN order and the works 
\cite{Jaranowski:1996nv,Konigsdorffer:2003ue,Nissanke:2004er} 
derived the corrections at that order using the HPF. 
See \cite{Blanchet:2006zz} for a review.

Now what is our method to achieve a point particle limit? Futamase 
\cite{Futamase87} proposed 
the strong field point particle limit. In this limit the binary star is 
first described as an extended object and then its radius is taken to
zero in a specific manner which will be explained in the Section II below.  
By this limit we obtain a point particle with strong internal gravity 
that is appropriate as a model of a relativistic compact star. 
With this limit and the method mentioned below we have successfully derived the 3
PN equations of motion that contains no undetermined coefficient 
\cite{Itoh:2003fy,Itoh:2003fz} and confirmed the earlier result
\cite{Damour:2001bu}.

The second aspect of our method that is different from the others is 
the way to derive equations of motion. In our method, using the local energy momentum conservation
law, we compute the force acting on the star by 
computing the gravitational energy momentum flux going through a suitably
defined surface around the star. This idea was used by Einstein, Infeld
and Hoffmann in their derivation of the 1 PN equations of motion 
\cite{Einstein:1938yz,Einstein:1940mt}. On the other hand, it was assumed in
some works (\cite{BFP98} at the 2.5 PN order,  
\cite{Blanchet:2000nv,Blanchet:2000ub} at the 3 PN order, and 
\cite{PW02,Nissanke:2004er} at the 3.5 PN order) that the binary star 
follows a ``geodesic'' of spacetime. However for an equal mass binary, 
the metric components diverge at the   
position of the star when it is described as a point particle. 
In those works \cite{BFP98,Blanchet:2000nv,Blanchet:2000ub,Nissanke:2004er},  
the divergence was regularized by 
the HPF (at the 2.5 and 3.5 PN order), or 
the combination of the HPF and the dimensional regularization (at the 3
PN order).  Our results up to the 3 PN order inclusively have 
fully agreed with these results. This agreement 
has confirmed that the star follows the regularized geodesics at least
through the 3 PN order. One of the
aims of the current paper is to extend it to the 3.5 PN order.  
(In this respect, Fukumoto et al \cite{Fukumoto:2006gv}  
showed that the star should follows  
the geodesic of the ``smooth part of the full metric'' 
where the smoothing is done by means of the 
surface integral approach. However, this work showed it for the case
of an extreme mass ratio binary.)

This paper is organized as follows. The next section explains 
our method of deriving the equations of motion, sketching each steps in
our method with emphases on 
the strong field point particle limit, the surface integral approach and
the scalings of the stress energy tensor of the matter assumed in our approach.
Sec. \ref{chap:formulation} explains how to derive 
the PN gravitational field and the PN equations of motion. 
After showing the formal structure of the 3.5 PN equations of motion 
in Sec. \ref{sec:structure},  
the 3.5 PN gravitational field in harmonic coordinates is derived in Sec. \ref{sec:field}.
Since we shall not use a Dirac delta functional nor assume 
any specific form of a momentum 
velocity relation, we have to derive the relation between the star's mass and its energy and
the one between the star's three momentum and its three velocity. 
Those relations at the 3.5 PN order are reported in the sections \ref{sec:mass-energy} and 
\ref{sec:momentum-velocity}. We show the resulting 3.5 PN equations of motion 
in Sec. \ref{sec:eom}.

Throughout this paper, we will use 
harmonic coordinates and the unit of $c=1=G$ where $c$ is the 
velocity of light and $G$ is the Newton's gravitational constant.  
A tensor with alphabetical indexes such as $x^i$,  
denotes a Euclidean three-vector. 
We raise or lower its indexes with a Kronecker
delta. For an object with Greek indexes such as $x^{\mu}$, its indexes are raised or lowered with a flat Minkowskian metric.
We shall call our previous papers \cite{IFA00,IFA01,Itoh:2003fz} Paper
I, II, and III, respectively.

\section{Key ideas in our derivation}
\label{sec:PNA}

In this section we briefly review our method to obtain post-Newtonian
equations of motion, listing the procedures in our method  
with emphases on the strong field point particle
limit, the surface integral approach and the  
associated scalings of the stress energy tensor of the matter. See 
\cite{IFA01,Futamase87,IFA00,Schutz85,AsadaFutamase97,Schutz80,FS83,Futamase83} 
for more details.

We first introduce an 
adimensional parameter $\epsilon$ which represents  
the slowness of the star's 
orbital velocity $\tilde v^i_{{\rm orb}}$.  
\begin{eqnarray*}
&&
\tilde{v}^i_{{\rm orb}} 
\equiv \frac{d z^i}{d t} \equiv \epsilon \frac{d z^i}{d \tau},  
\end{eqnarray*} 
where $z^i$ is the star's representative point and 
we assume $v^i \equiv d x^i/d \tau$ of order unity.  
The time coordinate $\tau$ is called the {\it Newtonian dynamical time} 
\cite{Futamase87,AsadaFutamase97}. The post-Newtonian scaling implies
that the square modulus of the orbital velocity is of order of the
inter-body gravity $\tilde m/r$ where $\tilde m$ and $r$ are the star's mass and the 
orbital radius, respectively. In terms of $\epsilon$, this means that 
the mass of the star $\tilde m$ is of order of $\epsilon^2$ \cite{IFA00,IFA01,Itoh:2003fz}. 
In what follows, $m$
(without a tilde) denotes the star's mass that is of order unity. The
smallness of the orbital velocity and the inter-body gravity are
represented in terms of $\epsilon$. Thus, $\epsilon$ is a post-Newtonian 
expansion parameter.  Henceforth we call the $(\tau,x^i)$
coordinate the near zone coordinate. 

The typical magnitude of the velocity of the system far away from the 
binary is the velocity of the gravitational wave emitted by the 
binary, or the velocity of light $c$. Hence there must be different coordinates 
appropriate in the far zone \cite{FS85,Futamase85}. 
The near zone coordinate is useful within, say, one wavelength of 
the gravitational waves (of order $\sim c \times$ 
(orbital period of the binary)) from the binary system center of the mass. 

Now, the next subsection sketches the principal steps of our derivation of
post-Newtonian equations of motion.

\subsection{Principal steps of our derivation}

Suppose we have the $n$ PN metric and the $n$ PN equations of motion.
To derive the $n+1$ PN equations of motion, we first solve the Einstein equations to the $n+1$ PN order 
and then compute the gravitational force acting on the binary star.
When solving the Einstein equations, we use the following techniques.
\begin{itemize}
\item Gauge: We use harmonic coordinates to solve the Einstein equations.
\item Einstein equations: We solve the harmonically relaxed Einstein
      equations where we use the Minkowskian wave operator for
      computational simplicity and where we rewrite the Einstein
      equations as  
      a set of retarded integrals in the same manner as in \cite{AndersonDecanio1975}. 
\item Boundary condition at infinity: We take the no-incoming radiation
      condition at the Minkowskian past null-infinity \cite{Fock1959,Damour1983}. 
\end{itemize}
Other previous works \cite{PW02,Nissanke:2004er} take essentially
the same approaches on all of the above points. 

Now it is well-known that the post-Newtonian approximation breaks down 
far away from the binary at high post-Newtonian orders (e.g., \cite{Ehlers1976}). 
To obtain the metric, it is thus convenient 
to divide the spacetime into the near zone and the far zone. 
The near
zone is a time-like world tube surrounding the binary stars. 
The center of the near zone is defined (if necessary) as the center of 
the mass of
the binary. 
The intersection between the $\tau = $constant spatial hypersurface and 
the world tube becomes a spatial 3-sphere. The radius of the sphere is
of order of the largest wavelength of the gravitational wave emitted by
the binary due to its orbital motion. Its mathematical definition shall
be given later in Sec. \ref{sec:FieldEq}. 
Then the far zone is the outside of the near zone. The domain of the
integration of the retarded integrals is now divided into that in the
near zone and that in the far zone. We compute two contributions
separately.  

In the near zone we can use the post-Newtonian approximation safely. 
We use the following method to obtain the near zone contribution.
\begin{itemize}
\item Near zone contribution: 
To obtain the near zone metric, we first expand the retarded integrals in
$\epsilon$ and change the domain of the integration from the Minkowskian null-cone to the $\tau = $ constant spatial
hypersurface. This hypersurface is denoted as $NZ$. 
\item Body zone: The region $NZ$ is divided into three regions. Two
      of the three are spheres called the body zones each of which
      surrounds each of the binary star. The remaining is the region outside of 
      the two body zones.
\item Multipole expansion: In each body zone, the integrals are
      evaluated by a multipole expansion. 
\end{itemize} 

In the multipole expansion, the multipole moments are defined as the 
volume integrals on the $\tau = $ constant spatial hypersurface 
whose integrands include the matter's stress energy 
tensor plus the gravitational stress energy tensor. In the usual 
post-Newtonian approximation, one assumes that the gravitational 
field is everywhere weak. Since the gravitational wave detectors 
search for relativistic compact binaries that have strong internal 
gravity, we would like to have a method where we use the post-Newtonian 
approximation only outside of the stars. One procedure which we adopt
here to solve this problem is proposed by Futamase \cite{Futamase87}.
\begin{itemize} 
\item Strong field point particle limit \cite{Futamase87}: 
With this limit the star's internal gravity can be assumed to be
      strong. Namely if the companion 
      star were absent, the star's mass would become the ADM mass. 
      We shall explain this limit in more detail in the next
      subsection. 
\end{itemize}
As in the Newtonian dynamics, the multipole moments are defined 
with respect to some reference point \cite{Dixon1979}. 
\begin{itemize}
\item Star's representative point: The representative point of the star
      (e.g., the center of the mass of the star) is
      defined by setting the star's dipole moment appropriately 
      in the same manner as in the Newtonian dynamics.  The star's
      multipole moments are defined with respect to this representative point.
\end{itemize}

As is mentioned in the introduction, this paper studies an intrinsically
spherically symmetric star. As the multipole moments above are defined as volume integrals on the 
$\tau = $ constant spatial hypersurface, we cannot obtain a spherically
symmetric star by simply making the multipole moments vanish.
This is because the Lorentz contraction makes such a star acquire apparent 
multipole moments when an observer sees the star moving with  
respect to the observer. We define the star's {\it intrinsic} multipole
moments in the generalized Fermi normal coordinate \cite{AB86} and set those zero to
obtain an intrinsically spherically symmetric star.

Now, in the far zone, it is well-known that a simple post-Newtonian 
expansion gives divergent 
integrals (e.g., \cite{Ehlers1976}). To solve this problem,  we use the following method. 
\begin{itemize}
\item Far zone contribution: we use the 
{\it Direct Integration of the Relaxed Einstein equations}  
(DIRE) method to evaluate the far zone contribution. 
\end{itemize}
This method was proposed by Will and his collaborators \cite{PW02,WW96,PW00}.
This method assumes that the binary is sufficiently stationary in
the long past. 
On the other hand, the work \cite{Nissanke:2004er} used the multipolar
post-Minkowskian (MPM) approach \cite{BlanchetDamour86} to obtain the
far zone metric. The asymptotic matching is done between the PN inner metric and the MPM 
outer metric \cite{PoujadeBlanchet2002}.

Having the metric components to the $n+1$ PN order, 
we now derive the $n+1$ PN equations of motion for relativistic compact binaries.
The following steps are used only in our method.
\begin{itemize}
\item Surface integral approach: Using the local conservation law of the stress 
      energy, we obtain expressions for the time derivative of the four 
      momentum in terms of the surface integrals over the star's 
      body zone surface.
      The integrands shall be the Landau-Lifshitz pseudo-tensor.  
      Thus, the surface integrals give the net flux going through the
      star's body zone surface and amount to the gravitational force acting on
      the star.
\item Four momentum - mass and velocity relation: As is written in the
      previous point, we compute the time derivative of the four
      momentum. To obtain equations of motion, we need to compute the time derivative of the 
      coordinate velocity. In our paper the star's four
      momentum is defined as the volume integral of the sum of the 
      matter's and gravitational stress energy tensors. It is
      non-trivial that the four momentum is proportional to the four
      velocity \cite{Dixon1979}. We develop 
      a method to derive the expression for the four momentum in terms 
      of the stars' masses and their coordinate velocities.
\item Gravitational energy momentum flux: Using the metric 
      components (the solution of the Einstein equations to the $n+1$ PN order), we evaluate the Landau-Lifshitz
      pseudo-tensor to the $n+1$ PN order.  
\item Surface integrals: We evaluate the surface integrals to the $n+1$ 
      PN order and obtain the $n+1$ PN equations of motion.
\end{itemize}
The previous works \cite{PW00,PW02} use the volume integral
approach where they assume a geodesic, multiply it by the conserved 
baryon density, and integrate it over the star on the time constant
hypersurface. Nissanke and Blanchet 
\cite{Nissanke:2004er} also assume a geodesic, but in their method 
they substitute the metric components into the geodesic and regularize
divergence due to their use of Dirac delta functionals.

In the following subsections, we shall explain in more detail the strong field point particle
limit, the surface integral approach, and the associated scalings 
of the stress energy tensor components. 
Other features listed above shall be explained in 
Sec. \ref{chap:formulation}. 
We apply the procedures listed in this 
section to the derivation of the 3.5 PN order equations of motion   
starting from Sec. \ref{sec:structure}.

\subsection{Strong Field Point Particle Limit}

One would describe a star as a point particle by making the radius of the
star zero. However, by this procedure, one would obtain 
a black hole rather than a point particle. The radius of the star cannot be made smaller than of order
of its mass (in the unit of $G = 1 = c$). Futamase \cite{Futamase87} proposed 
that we may obtain a point particle model for the star by taking a limit
where both the radius of the star and its mass shrink at the same rate. 
Since the post-Newtonian scaling implies that the star's mass $\tilde m$ 
is $O(\epsilon^2)$, we assume that its radius is also $O(\epsilon^2)$. 
The point particle limit is now achieved in the limit where $\epsilon$ goes
to zero.

The usual post-Newtonian approximation assumes the gravitational 
field is everywhere weak even inside the stars. 
On the other hand, with Futamase's procedures we obtain a point particle with finite internal 
gravity, since a typical magnitude of the self-gravitational field (the mass
over the radius) of the point particle achieved by this procedure 
is finite irrespective of $\epsilon$. 
For this reason, this limit is called the {\it strong field point particle limit}.
Note that the inter-body gravity (the mass over the orbital separation) 
is $O(\epsilon^2)$ and the post-Newtonian approximation applies for the
aorbital motion.  With this limit, we can derive post-Newtonian equations of motion 
for a binary star whose component stars have strong internal gravity.

\subsection{Surface Integral Approach and Body Zone}
\label{sec:EIHApproach}

One way to derive equations of motion of a star is a surface integral approach 
where we compute the total gravitational energy momentum flux going through 
a surface around the star. When we let the radius of the surface 
shrink to zero and when 
the radius of the star goes to zero faster than that of the surface in the 
point particle limit, we would then obtain equations of motion 
for the star as a point particle. For this purpose, we introduce two body 
zones $B_A$ for the stars $A=1,2$ as 
$B_A \equiv\{x^i | |\vec{x} - \vec{z}_A(\tau)| < \epsilon R_A \}$.  
Here $z_A^i(\tau)$ is 
a representative point of the star $A$, e.g., the center of the mass
of the star $A$. We shall later fix $z_A^i$ by 
specifying the star's dipole moment as in the Newtonian dynamics.
The body zone radius $\epsilon R_A$ 
are much smaller than the orbital separation but 
is larger than the radius of the star for any $\epsilon$ 
(Recall that the radius of the star decreases proportionally 
to $\epsilon^2$ in the 
strong field point particle limit while the body zone radius does so 
proportionally to $\epsilon$). 
Note that the two body zones do not overlap each other. 
Finally, $R_A$ are constant, i.e., $d R_A/d\tau =0$. 
Other than these conditions, $R_A$ are arbitrary.

Now the above scalings of the masses and the radii of the stars motivate 
us to introduce a body zone coordinate for the star $A$ as 
$(\tau, \alpha_A^{\underline{i}})$ where $\alpha_A^{\underline{i}} \equiv \epsilon^{-2} (x^i -z_A^i(\tau))$.
The scalings of the body zones and the body zone coordinates 
give us a situation where in the body zone coordinate $A$ 
the star $A$ does not shrink as $\epsilon \rightarrow 0$  
while the boundary of the body zone expands to infinity. 
Thus, it is appropriate to define
the star's characteristic quantities such as its mass using 
the body zone coordinate. Moreover, 
since the body zone boundary $\pa B_A$ is far away from the surface
of the star $A$ (in its body zone coordinate), 
we can evaluate 
explicitly the gravitational energy momentum flux
on $\pa B_A$ using the post-Newtonian gravitational field. 
After evaluating the surface integrals, 
we make the body zone shrink to derive the equations of 
motion for the compact star. 

Possible effects of the internal structures of the compact stars 
are coded in the multipole moments of the stars. 
These moments in turn 
appear in the gravitational energy momentum flux and 
would affect the orbital motion. However, in this paper we shall concentrate on  
spherically symmetric stars and ignore those multipole moments.

\subsection{Scalings of the Matter Stress Energy Tensor}
\label{sec:Scaling}

The scalings of the radii and the masses of the stars 
indicate that the matter density is $O(\epsilon^{-4})$ 
in the $(t,x^i)$ coordinate (or 
$\epsilon^{-2}$ in the near zone coordinate $(\tau,x^i)$). 
We further assume that 
the internal time scale of the star is comparable to 
that of the binary orbital motion and is $O(\epsilon)$. Namely,  
in this paper we assume that the star is pressure supported and non-rotating, 
although an extension to rapidly rotating stars is straightforward 
\cite{Futamase87}. 

In terms of the stress energy tensor of the matter 
$T^{\mu\nu}$ 
(or 
the source terms of the harmonically 
relaxed Einstein equations $\Lambda^{\mu\nu}$ in
Eq. (\ref{eq:RelaxedEinsteinEquations}) below), these scalings 
imply 
$T^{\tau\tau} = O(\epsilon^{-2})$,
$T^{\tau \underline{i}} = O(\epsilon^{-4})$,    
$T^{\underline{i}\underline{j}} = O(\epsilon^{-8})$ 
where the underlined indexes mean that for any tensor $S^i$,
$S^{\underline i} = \epsilon^{-2} S^i$ and reminds the scaling of the
body zone spatial coordinate \cite{Futamase87}. 
Because a star moves in a gravitational field, the tensor 
components in the body zone coordinate are different from 
those in the near zone coordinate. Let us denote the matter's stress energy
tensor in the body zone coordinate $(\tau,\alpha_A^i)$ by $T_A^{\mu\nu}$ and that in the
near zone coordinate $(\tau,x^i = z_A^i +
\epsilon^{-2}\alpha_A^{\underline i})$ by
$T_{NZ}^{\mu\nu}$. Transforming $T_A^{\mu\nu}$ to the near zone coordinate
we obtain \cite{Futamase87}
\begin{eqnarray*}
T_{NZ}^{\tau\tau} &=& T_A^{\tau\tau},
\nonumber \\
T_{NZ}^{\tau i} &=& \epsilon^2 T_A^{\tau \underline i} + v_A^i T_A^{\tau\tau},
\nonumber \\ 
T_{NZ}^{i j} &=& \epsilon^{4}T_A^{\underline{ij}} +
 2\epsilon^2v_A^{(i}T_A^{\underline j)\tau} + v_A^iv_A^j T_A^{\tau\tau}.  
\end{eqnarray*}
Hence the stress energy tensor components of the matter in the near zone 
coordinate varies with respect to $\epsilon$ 
in the same way as those in the body zone coordinate,  
or in short, 
$T_{NZ}^{\tau\tau} = O(\epsilon^{-2})$,
$T_{NZ}^{\tau \underline{i}} = O(\epsilon^{-4})$,    
$T_{NZ}^{\underline{i}\underline{j}} = O(\epsilon^{-8})$ \cite{Itoh:2003fz}.

\section{Mathematical Formulation}
\label{chap:formulation}

Based on the idea explained in the previous section, this section formulates our 
method to derive the equations of motion. We first explain how to solve 
the Einstein equations, then show how we achieve the surface integral 
approach.

\subsection{Field Equation}
\label{sec:FieldEq}

In the surface integral approach, we need to compute the gravitational field 
near the body zone boundary where the field is well described
by the post-Newtonian approximation and slightly deviates from the flat
metric. We define a deviation field $h^{\mu\nu}$ as 
\begin{eqnarray}
&&
h^{\mu\nu} \equiv  \eta^{\mu\nu} - \sqrt{-g}g^{\mu\nu},
\end{eqnarray}
where $\eta^{\mu\nu} = {\rm diag}{(-\epsilon^2,1,1,1)}$ is the flat
metric in the near zone coordinate $(\tau,x^i)$ and  $g$ is the determinant of the metric. 
The indexes of $h^{\mu\nu}$ 
are raised or lowered  by the flat metric.

Now we impose a harmonic coordinate condition 
$
 h^{\mu\nu}{}_{,\nu}=0  
$
where the comma denotes a  partial derivative. 
In the harmonic gauge, we can recast the Einstein 
equations into a relaxed form, 
\begin{eqnarray}
\Box h^{\mu\nu} = -16\pi \Lambda^{\mu\nu}, 
\label{eq:RelaxedEinsteinEquations}
\end{eqnarray}
 where $
\Box = \eta^{\mu\nu}\pa_{\mu}\pa_\nu$
is the flat spacetime d'Alembertian. 
The source term $\Lambda^{\mu\nu}$ of the relaxed Einstein
equations consists of two pseudo-tensors. The first is the sum of the
stress energy tensor of the stars denoted by $T^{\mu\nu}$ and  
the Landau-Lifshitz pseudo-tensor $t^{\mu\nu}_{LL}$ \cite{LL1975}. The
second arises due to our use of the flat spacetime  d'Alembertian
instead of the curved spacetime one. The explicit expressions are 
\begin{eqnarray}
&&
\Lambda^{\mu\nu} \equiv \Theta^{\mu\nu}
+\chi^{\mu\nu\alpha\beta}{}_{,\alpha\beta} ,
\label{DefOfLambda} \\ 
&&
\Theta^{\mu\nu} \equiv (-g) (T^{\mu\nu}+t^{\mu\nu}_{LL}) , \\
&&
\chi^{\mu\nu\alpha\beta} \equiv \frac{1}{16\pi} 
(h^{\alpha\nu}h^{\beta\mu}
-h^{\alpha\beta}h^{\mu\nu}) . 
\label{eq:DefOfChi} 
\end{eqnarray}
The harmonic condition on $h^{\mu\nu}$ implies a local   
energy momentum conservation law;  
\begin{equation}
\Lambda^{\mu\nu}{}_{,\nu}=0 . 
\label{conservation}
\end{equation}
Note that $\chi^{\mu\nu\alpha\beta}\mbox{}_{,\alpha\beta}$ 
itself is divergence free, i.e.,
$\chi^{\mu\nu\alpha\beta}\mbox{}_{,\alpha\beta\nu} = 0$.

We can formally rewrite the relaxed Einstein equations as retarded
integrals;  
\begin{equation}
h^{\mu\nu}(\tau,x^i)=4 \int_{C(\tau, x^k)} d^3y 
\frac{\Lambda^{\mu\nu}(\tau-\epsilon|\vec x-\vec y|, y^k; \epsilon)} 
{|\vec x-\vec y|}, 
\label{IntegratedREE}
\end{equation}
where $C(\tau, x^k)$ means the past light cone emanating 
{}from the event $(\tau, x^k)$. 
We have assumed no homogeneous solution
of the relaxed Einstein equations. 
It is well-known that 
this condition can be deduced from 
the Minkowskian no-incoming radiation condition 
(See, e.g., the section 92 of \cite{Fock1959} or the section 6 of \cite{Damour1983}).

We solve the Einstein equations as follows. First we split the
domain of the integration into two zones: the near zone and the far zone. 
The near zone is  
the neighborhood of the gravitational wave source where the wave 
character of the gravitational radiation is not manifest. 
In this paper, as in our previous papers, 
we define the near
zone as a time-like world tube surrounding the binary stars. 
The center of the near zone is defined (if necessary) as the center of
the mass of the binary. 
The intersection between the $\tau = $constant spatial hypersurface and 
the world tube becomes a spatial 3-sphere. Mathematically, 
denoting the harmonic coordinate distance from the center by
$r$,  the near zone is defined as $r < {\cal R}/\epsilon$ where 
${\cal R}/\epsilon$ is of order of the largest wavelength of the
gravitational wave emitted by the binary due to its orbital motion. 
We assume ${\cal R}/\epsilon$ sufficiently large so that the near zone
 covers the binary stars. Finally, ${\cal R}$ is constant in time, i.e.,
 $d {\cal R}/d\tau = 0$. 
Otherwise ${\cal R}$ is arbitrary.
The $\epsilon^{-1}$ scaling of the near zone radius is derived from the 
$\epsilon$ dependence of the wavelength of the gravitational wave
emitted by the binary. 
The outside of  
the  near zone is the far zone where the retardation effect of the 
field is crucial. 

For the retarded integrals in the far zone, we evaluate it  
using the 
{\it 
Direct Integration of the Relaxed Einstein equations}  
(DIRE) method. This method was proposed by 
Will and his collaborators 
\cite{PW02,WW96,PW00}. DIRE directly and nicely fits into our 
formalism since it utilizes the relaxed Einstein equations 
in the harmonic gauge. Using DIRE, 
Pati and Will \cite{PW00} showed that 
the far zone contribution to the near zone field 
affects (physically) the orbital motion 
starting at the 4 PN order In fact, 
this result was first obtained by Blanchet and Damour \cite{BD88}. 
Although we do not show our explicit computation in this paper,  
we have followed the DIRE method and 
checked that the far zone contribution does not affect the 
equation of motion through the 3.5 PN order.
We shall thus focus our attention on the 
near zone contribution to the near zone field and neglect 
all the far zone contributions to the near zone field.

\subsection{Near Zone Contribution}

For the near zone contribution, we first 
make retardation expansion 
and change the domain of the integration to 
a $\tau =$ constant spatial hypersurface 
\begin{equation}
h^{\mu\nu}(\tau,x^i)=4
\sum_{n=0} 
\frac{(-\epsilon)^n}{n!}\left(\frac{\pa}{\pa \tau}\right)^n
\int_{NZ} d^3y 
|\vec x-\vec y|^{n-1}
\Lambda_{NZ}^{\mu\nu}(\tau, y^k; \epsilon).
\label{RetardedEIREE}
\end{equation}
where $NZ$ denotes the near zone and we attach the subscript 
$NZ$ to $\Lambda^{\mu\nu}$ to clarify that they are quantities 
in the near zone.
Note that the above integral depends on the arbitrary length
${\cal R}$ in general. The cancellation between the ${\cal R}$ 
dependent terms in the far zone contribution and those in the near zone 
contribution through all 
the post-Newtonian order was shown by Pati and Will \cite{PW00}.  
This paper uses their method, and 
hence we safely neglect all the ${\cal R}$ dependent terms other 
than those where ${\cal R}$  appears in arguments of logarithms. 
This is to make the arguments of all possible logarithmic terms 
adimensional.

Second we split
the integral into two parts; contribution from the body zone
$B= B_1 \cup B_2$, 
and from elsewhere, $NZ/B$. We thus evaluate
the following two types of integrals 
\begin{eqnarray}
&&
h^{\mu\nu} = \sum_{n=0}
\left(h^{\mu\nu}_{Bn} + h^{\mu\nu}_{NZ/Bn}\right), 
\label{TotFieldRetExpandAbst} 
\\
&&
h^{\mu\nu}_{Bn} = 
4
\frac{(-\epsilon)^n}{n!}\left(\frac{\pa}{\pa \tau}\right)^n
\epsilon^6 \sum_{A=1,2}\int_{B_A}
 d^3\alpha_A 
\frac{\Lambda_{NZ}^{\mu\nu}(\tau,\vec z_A + \epsilon^2\vec{\alpha}_A)}
{|\vec r_A-\epsilon^2\vec{\alpha}_A|^{1-n}} ,  
\label{eq:BConstributionSC} 
\\
 &&
h^{\mu\nu}_{NZ/Bn} = 
4
\frac{(-\epsilon)^n}{n!}\left(\frac{\pa}{\pa \tau}\right)^n
\int_{NZ/B}  d^3y 
\frac{\Lambda_{NZ}^{\mu\nu}(\tau,\vec{y})}
{|\vec x-\vec y|^{1-n}},
\label{NBContributionSC}
 \end{eqnarray}
where $\vec r_A \equiv \vec x - \vec z_A$. 
We shall deal with these two contributions successively 
in the followings.

\subsubsection{Body Zone Contribution}

As for the body zone contribution,  
we make  multipole expansion using 
the scaling of the integrand ($\Lambda_{NZ}^{\mu\nu}$) in the body zone. 
For example, the $n=0$ part in  Eq. (\ref{eq:BConstributionSC}), 
$h_{B n=0}^{\mu\nu}$, gives
\begin{eqnarray}
h^{\tau\tau}_{B n=0} &=& 4 \epsilon^4 \sum_{A=1,2}
\left(\frac{P_A^{\tau}}{r_A} + \epsilon^2 \frac{D_A^k r^k_A}{r_A^3} +
 \epsilon^4 \frac{3 I_A^{<kl>} r^k_A r^l_A}{2 r_A^5} +
 \epsilon^6 \frac{5 I_A^{<klm>} r^k_A r^l_A r_A^m}{2 r_A^7} 
\right)
\nonumber \\
\mbox{} &&+ O(\epsilon^{12}),
\label{hBtt} \\ 
h^{\tau i}_{B n=0} &=& 4 \epsilon^4 \sum_{A=1,2}
\left(\frac{P_A^{i}}{r_A} + \epsilon^2 \frac{J_A^{ki} r^k_A}{r_A^3} 
+ \epsilon^4 \frac{3 J_A^{<kl>i} r^k_A r^l_A}{2 r_A^5} 
 \right)
 + O(\epsilon^{10}),
\label{hBti} \\ 
h^{ij}_{B n=0} &=& 4 \epsilon^2 \sum_{A=1,2}
\left(\frac{Z_A^{ij}}{r_A} + \epsilon^2 \frac{Z_A^{kij} r^k_A}{r_A^3} +
 \epsilon^4 \frac{3 Z_A^{<kl>ij} r^k_A r^l_A}{2 r_A^5} 
+
 \epsilon^6 \frac{5 Z_A^{<klm>ij} r^k_A r^l_A r^m_A}{2 r_A^7} 
\right)
\nonumber \\
\mbox{} &&+ O(\epsilon^{10}), 
 \label{hBij}
\end{eqnarray}
where $ r_A \equiv |\vec{r}_A|$.
The quantity with $<>$ 
denotes a symmetric and tracefree (STF) operation on the indexes  
between the brackets. 
To derive the 3.5 PN equations of motion,  we need  
$h^{\tau\tau}$ up to $O(\epsilon^{11})$ and 
$h^{\mu i}$  up to $O(\epsilon^9)$.

In the above equations we defined the multipole moments of the star $A$ 
as 
\begin{eqnarray}
&&
I_A^{K_l} \equiv \epsilon^2 \int_{B_A}
d^3\alpha_A\, \Lambda^{\tau \tau}_{NZ} \alpha_A^{\underline{K_l}}, 
\label{eq:NZCTTmoment}
\\   
&&
J_A^{K_li} \equiv \epsilon^4 \int_{B_A}
d^3\alpha_A\, \Lambda^{\tau \underline i}_{NZ} \alpha_A^{\underline{K_l}},  
\label{eq:NZCTImoment}
\\
&&
 Z_A^{K_lij} \equiv \epsilon^8 \int_{B_A}
d^3\alpha_A\, \Lambda^{\underline{i}\underline{j}}_{NZ} 
\alpha_A^{\underline{K_l}},  
\label{eq:NZCIJmoment}
\end{eqnarray}
where the capital index denotes a set of collective indexes, 
$I_l \equiv i_1 i_2 \cdot\cdot\cdot i_l$  and
$\alpha_A^{\underline{I_l}} \equiv
\alpha_A^{\underline i_1}\alpha_A^{\underline i_2}
\cdot\cdot\cdot\alpha_A^{\underline i_l}$.  
Then 
$P_A^{\tau} \equiv I^{I_0}_A$, $D_A^{i_1} \equiv I^{I_1}_A$,
$P_A^{i_1} \equiv J^{I_1}_A $.
We simply call $P^{\mu}_A$ the four momentum of the star $A$,
$P^{i}_A$ the (three) momentum, and $P^{\tau}_A$ the energy. 
Also we call $D^{i}_A$ the dipole moment and
$I^{ij}_A$ the quadrupole moment.

Then we transform these moments into more convenient forms 
using the conservation law Eq. (\ref{conservation}). 
In the following, $v_A^i \equiv \dot{z}^i_A$, an overdot 
denotes a $\tau$ time
derivative, and $\vec y_A \equiv \vec y - \vec z_A$. 
Noticing that the body zone radii are constant 
, i.e., $\dot R_A = 0$, we have 
\begin{eqnarray}
&&
P^{i}_A = P^{\tau}_A v^i_A + Q_A^i +
\epsilon^2 \frac{d D_A^i}{d\tau},    
\label{MomVelRelation} \\ 
&&
J^{ij}_A =
\frac{1}{2}
\left(M_A^{ij} + \epsilon^2 \frac{d I_A^{ij}}{d\tau}\right) 
+ v_A^{(i} D_A^{j)} + \frac{1}{2}\epsilon^{-2} Q_A^{ij},
\label{JijToMij}
\\
Z^{ij}_A &=& \epsilon^2 P^{\tau}_A v_A^i v_A^j + 
 \frac{1}{2}\epsilon^6\frac{d^2 I_A^{ij}}{d\tau^2} +
2\epsilon^4 v_A^{(i}\frac{d D_A^{j)}}{d\tau} +
\epsilon^4 \frac{d v_A^{(i}}{d\tau}D_A^{j)}  \nonumber \\
\mbox{} &+& 
\epsilon^2 Q_A^{(i}v_A^{j)} + \epsilon^2 R_A^{(ij)} +
\frac{1}{2}\epsilon^2\frac{d Q_A^{ij}}{d\tau},
\label{StrVelRelation}
\\
&&
Z^{kij}_A = \frac{3}{2}A_A^{kij} - A_A^{(ij)k},
\label{ZkijToAkij}
\end{eqnarray}
where 
\begin{eqnarray}
&&
M_A^{ij} \equiv 2\epsilon^4\int_{B_A}d^3\alpha_A
\alpha_A^{[\underline i}\Lambda_{NZ}^{\underline j]\tau} , \\
&&
Q_A^{K_li} \equiv \epsilon^{-4}
 \oint_{\pa B_A} dS_m
 \left(\Lambda^{\tau m}_{NZ}  - v_A^m 
\Lambda^{\tau\tau}_{NZ} \right) y_A^{K_l} y_A^i
\label{QL} , \\ 
&&
R_A^{K_lij} \equiv \epsilon^{-4}
 \oint_{\pa B_A} dS_m
 \left(\Lambda^{m j}_{NZ}  - v_A^m 
\Lambda^{\tau j}_{NZ} \right) y_A^{K_l} y_A^i,
\label{RL} 
\end{eqnarray}
and
\begin{eqnarray}
&&
A_A^{kij} \equiv \epsilon^2 J_A^{k(i}v_A^{j)} + \epsilon^2 v_A^k J_A^{(ij)}
+R_A^{k(ij)} + \epsilon^4 \frac{dJ_A^{k(ij)}}{d\tau}.
\end{eqnarray}
The symbol 
$[$ $]$ (or $($ $))$ attached to some of the indexes 
denotes anti-symmetrization (or  symmetrization) on the indexes 
between the brackets.
$M_A^{ij}$ is the spin of the star $A$ which is set to be zero for the
purpose of the current paper.   
Eq. (\ref{MomVelRelation}) gives a  momentum-velocity relation. 
Thus our momentum-velocity  
relation is a direct analogue of the Newtonian momentum-velocity
relation \cite{Dixon1979}.  In general, we have
\begin{eqnarray}
&&
J_A^{K_li} = J_A^{(K_li)} +  \frac{2l}{l+1}J_A^{(K_{l-1}[k_l)i]}, \\
&&
Z_A^{K_lij} = \frac{1}{2}
\left[Z_A^{(K_li)j} + \frac{2 l}{l+1}Z_A^{(K_{l-1}[k_l)i]j} +
Z_A^{(K_lj)i} + \frac{2 l}{l+1}Z_A^{(K_{l-1}[k_l)j]i} \right],
\end{eqnarray}
and 
\begin{eqnarray}
&&
J_A^{(K_li)} = \frac{1}{l+1}\epsilon^2\frac{d I_A^{K_li}}{d\tau}
+ v_A^{(i}I_A^{K_l)} + \frac{1}{l+1}\epsilon^{-2l}Q_A^{K_li}, 
 \label{JLiToILi}
\\
&&
Z_A^{(K_li)j} + Z_A^{(K_lj)i} =
\epsilon^2 v_A^{(i}J_A^{K_l)j} +
\epsilon^2 v_A^{(j}J_A^{K_l)i} +
\frac{2}{l+1}
\epsilon^4\frac{d J_A^{K_l(ij)}}{d\tau}
+ 
\frac{2}{l+1}
\epsilon^{-2l + 2}R_A^{K_l(ij)}.
\label{ZLijToJLij} 
\end{eqnarray}

Because the energy of the star, $P_A^{\tau}$, is not the mass of the star, we
shall find the relation between these two in
Sec. \ref{sec:mass-energy}. We may set the dipole moment $D_A^i$
to be zero to define the center of the mass of the star $z_A^i$, but we
will later choose a different value for $D_A^i$ for convenience. 
As for the multipole moments 
$I_A^{K_l}$, $J_A^{(K_{l-1}[k_l)i]}$ 
and $Z_A^{(K_{l-1}[k_l)i]j}$, we may equate those to zero, as it is our
aim to derive equations of motion for a spherically symmetric star.  
However, there is a subtlety here, which we will explain in the next
section.

\subsubsection{A Spherically Symmetric Star and the Star's Multipole Moments}
\label{sec:sphericalStar}

As mentioned in the introduction, this paper  
studies a binary consisting of 
two spherically symmetric compact stars. In other words,
all the multipole moments of the star defined in an  
appropriate reference coordinate 
where effects of its orbital motion and 
the companion star are removed (modulo, namely,
the tidal effect) vanish. 
We adopt the generalized Fermi normal
coordinate (GFC) \cite{AB86} as the reference coordinate.

We have defined the multipole moments of the star $A$ as Eqs. 
(\ref{eq:NZCTTmoment}), (\ref{eq:NZCTImoment}), and (\ref{eq:NZCIJmoment}). 
Those are defined on the $\tau$ = constant three-surface and 
differ from ones defined in the GFC.  Then a question specific to our formalism is if  
the differences between the multipole moments 
defined in Eqs. (\ref{eq:NZCTTmoment}), (\ref{eq:NZCTImoment}), 
and (\ref{eq:NZCIJmoment}) and 
the those in the GFC give purely monopole terms.  
Roughly speaking, a body that is spherical in its GFC may acquire 
apparent multipole moments due to, for example, the Lorentz contraction. 

At the 3 PN order, this problem was addressed in 
the appendix C of Paper III and  
the differences were mainly attributed to the shape of 
the body zone. The body zone $B_A$ which is spherical 
in the near zone coordinate (NZC) is not spherical in the GFC 
because of a kinematic effect (Lorentz contraction). 
In fact, the difference between the GFC quadrupole moment and the NZC one
was found to contain monopole terms. 

At the 3.5 PN order, no NZC multipole moment was found to 
contain purely monopole terms. We shall be back to this point later in
Sec. \ref{LorentzContraction}.  
We equate all those NZC multipoles 
$I_A^{K_l}$ ($l\ge 3$),
$J_A^{(K_{l-1}[k_l)i]}$ ($l\ge 1$) 
and $Z_A^{(K_{l-1}[k_l)i]j}$ ($l\ge 1$) to zero from now on in this paper.

\subsubsection{$NZ/B$ Contribution}

About the $NZ/B$ contribution, 
since the integrand 
$\Lambda_{NZ}^{\mu\nu} = -gt_{LL}^{\mu\nu} +
\chi^{\mu\nu\alpha\beta}\mbox{}_{,\alpha\beta}$ is at least 
quadratic in the small deviation field $h^{\mu\nu}$,
we make the post-Newtonian expansion in the integrand. 
Then, basically, with the help of a (super-)potential $g(\vec{x})$
which satisfies $\Delta g(\vec x) = f(\vec{x})$,
$\Delta = \partial_i\partial^i$ denoting Laplacian, we have for each integral 
(, e.g., $n=0$ term in Eq. (\ref{NBContributionSC}))    
\begin{eqnarray}
\int_{NZ/B} d^3y \frac{f(\vec{y})}{|\vec{x}-\vec{y}|}  &=&
- 4 \pi g(\vec{x}) + \oint_{\pa(NZ/B)}dS_k
\left[\frac{1}{|\vec{x}-\vec{y}|}
 \frac{\pa g(\vec{y})}{\pa y^k} - g(\vec{y})\frac{\pa}{\pa y^k}
 \left(\frac{1}{|\vec{x} - \vec{y}|}\right) \right].  
\label{NBcontribution}
\end{eqnarray} 
Eq. (\ref{NBcontribution}) can be proved without using 
a Dirac delta functional (see Appendix B of Paper III). 
For $n \ge 1$ terms in Eq. (\ref{NBContributionSC}), we use
appropriate (super-)potentials many times to convert all the volume integrals into surface
integrals and bulk terms (``$- 4 \pi g(\vec x)$'')  
\footnote{Notice that when solving a Poisson equation 
$\Delta g(\vec x) = f(\vec{x})$, a particular solution 
suffices for our purpose. By virtue of the surface integral 
term in Eq. (\ref{NBcontribution}), it is not necessary to take account
of a homogeneous solution of the Poisson equation.}.

Finding the super-potentials is one of the most formidable 
task especially when we proceed to a high post-Newtonian order.   
Fortunately, at the 3.5 PN order, all the required 
super-potentials are available (See \cite{PW02,Nissanke:2004er} and
Sec. \ref{sec:3hPNNBField} below).

\subsection{General Form of the Equations of Motion}

{}From the definition of the four momentum (Eq. (\ref{eq:NZCTTmoment}) with $l=0$) 
and the conservation law Eq. (\ref{conservation}) we obtain an 
evolution equation for the four momentum; 
\begin{equation}
\frac{dP_A^{\mu}}{d\tau} = -\epsilon^{-4}
 \oint_{\pa B_A} dS_k\, \Lambda^{k\mu}_{NZ}
+\epsilon^{-4}
v_A^k \oint_{\pa B_A} dS_k\, \Lambda^{\tau\mu}_{NZ}. 
\label{EvolOfFourMom_Lambda}
\end{equation}

We have defined the four momentum by a volume integral of
$\Lambda_{NZ}^{\tau\nu}$. Because 
$\chi^{\tau\nu\alpha\beta}\mbox{}_{,\alpha\beta} = \chi^{\tau\nu\alpha
i}\mbox{}_{,\alpha i}$, we can transform the volume integral of 
$\chi^{\tau\nu\alpha\beta}\mbox{}_{,\alpha\beta}$ into a surface
integral form and can evaluate it explicitly in terms of the mass,
velocity and the orbital separation. As a result, it is straightforward
to see that $\chi$ part of Eq. (\ref{EvolOfFourMom_Lambda}) is a trivial 
identity. The identity to the 3 PN order was shown in Appendix E of
Paper III by an explicit calculation. This observation implies that 
equations of motion can be derived from $\Theta$ part of
Eq. (\ref{EvolOfFourMom_Lambda}). We thus define the 
$\Theta$ part of the four momentum, 
dipole moment, and $Q_A^i$ integral as 
\begin{eqnarray}
&&
P_{A\Theta}^{\mu} \equiv \epsilon^2 \int_{B_A}
d^3\alpha_A\, \Theta^{\mu \tau}_{NZ}, 
\\
&&
D_{A\Theta}^i \equiv 
\epsilon^2 \int_{B_A}d^3\alpha_A\, 
\alpha_A^{\underline{i}}
\Theta_{NZ}^{\tau\tau},
\\  
&&
Q_{A\Theta}^i \equiv   \epsilon^{-4}
 \oint_{\pa B_A} dS_k
 \left(\Theta^{\tau k}_{NZ}  - v_A^k 
\Theta^{\tau\tau}_{NZ} \right) y_A^i.  
\end{eqnarray}  
Correspondingly, we split the momentum-velocity relation 
Eq. (\ref{MomVelRelation}) and 
the evolution equation for the four momentum 
Eq. (\ref{EvolOfFourMom_Lambda}) into the $\Theta$ part 
and the $\chi$ part. Note that the $\chi$ part of the 
multipole moments and the $Q_A^{K_li}$ and $R_A^{K_lij}$ integrals still 
in principle affect the equations of motion through the field, as 
$\chi^{\mu\nu\alpha\beta}\mbox{}_{,\alpha\beta}$  emerges as the
difference between the curved spacetime d'Alembertian and the flat 
spacetime one and thus should affect the gravitational field.
Indeed the $\chi$ part of the star's energy 
shown in Appendix \ref{sec:chipart} 
affects the acceleration at the 3.5 PN order. 

The $\Theta$ part of (\ref{EvolOfFourMom_Lambda}) is 
\begin{equation}
\frac{dP_{A\Theta}^{\mu}}{d\tau} = -\epsilon^{-4}
 \oint_{\pa B_A} dS_k\, \Theta^{k\mu}_{NZ}
+\epsilon^{-4}
v_A^k \oint_{\pa B_A} dS_k\, \Theta^{\tau\mu}_{NZ}. 
\label{EvolOfFourMom}
\end{equation}
Substituting  the $\Theta$ part of the momentum-velocity relation 
into the spatial components of  Eq. (\ref{EvolOfFourMom}), we obtain
the general form of the equations of motion for the star $A$;   
\begin{eqnarray}
P_{A\Theta}^{\tau}\frac{dv_A^i}{d\tau} &=&
 -\epsilon^{-4}
 \oint_{\pa B_A} dS_k\, \Theta^{ki}_{NZ}
+ \epsilon^{-4}
v_A^k \oint_{\pa B_A} dS_k\, \Theta^{\tau i}_{NZ} 
\nonumber\\
&&
 +\epsilon^{-4}
 v_A^i \left( \oint_{\pa B_A} dS_k\, \Theta^{k\tau}_{NZ}
-v_A^k \oint_{\pa B_A} dS_k\, \Theta^{\tau\tau}_{NZ} \right)
\nonumber\\
&&-\frac{dQ_{A\Theta}^i}{d\tau}  - \epsilon^2 \frac{d^2 D_{A\Theta}^i}{d \tau^2}.  
\label{generaleom}
\end{eqnarray} 

All the right hand side terms in Eq. (\ref{generaleom}) except for the
dipole moment are expressed as surface integrals. 
We can specify the
value of $D_{A\Theta}^i$ freely to determine the representative point
$z_A^i(\tau)$ of the star $A$.

In Eq. (\ref{generaleom}), $P_{A\Theta}^{\tau}$
rather than the mass of the star $A$ appears. Hence
we have 
to derive a relation between the mass and $P_{A\Theta}^{\tau}$.
We shall derive the relation by solving the temporal component of
the evolution equation (\ref{EvolOfFourMom}) 
functionally. In fact, 
at the lowest order, we have shown in Paper II that  
\begin{eqnarray}
\frac{dP_{A\Theta}^{\tau}}{d\tau} &=& O(\epsilon^2).
\label{Eq:LowestPtaudot}
\end{eqnarray}
Then we define the mass of the star $A$ as  
the integrating constant of this equation;   
\begin{eqnarray}
&& 
m_A \equiv \lim_{\epsilon \to 0}P_{A\Theta}^{\tau}.  
\label{DefOfMass}
\end{eqnarray}
$m_A$ is the ADM mass that the star $A$ had if the star 
$A$ were isolated.  
We took $\epsilon$ zero limit in Eq. (\ref{DefOfMass}) 
to ensure that the mass defined above  does not include the effect of the
companion star and the orbital motion of the star itself. 
Some subtleties
about this definition were discussed in Paper II.
By definition $m_A$ is constant. The procedure that we solve the 
evolution equation of $P_{A\Theta}^{\tau}$ and obtain the mass 
energy relation is achieved up to the 3.5 PN order successfully 
and the result will be shown in Sec. \ref{sec:mass-energy}.

\subsection{On the Arbitrary Constant $R_A$}
\label{Arbitrariness}

Our final remark in this section is on the two  
arbitrary constants $R_A$. 
Since we introduce the body zones by hand,
the arbitrary body zone radii $R_A$
seem to appear in the metric, the multipole moments of the stars,  
and the equations of motion. Paper II has proved that 
the surface integrals in the general equations of motion Eq. (\ref{generaleom}) 
do not depend on $R_A$ through any order of the post-Newtonian 
iteration. The appendix D of Paper III explained that 
the field and the multipole moments are independent of $\epsilon R_A$.

Practically, those two accounts justify that we safely discard   
all the $\epsilon R_A$ dependent terms except for 
logarithms of $\epsilon R_A$ that appear in the course of computations. 
We keep $\ln \epsilon R_A$ dependent terms to make 
the arguments of the logarithms adimensional.

We here emphasize that we discard the  $\epsilon R_A$ dependent 
terms in the field first 
and  then evaluate the surface integrals in the 
general form of the equations of motion using the field independent of 
$\epsilon R_A$. We then 
discard the $\epsilon R_A$ dependent terms  
arising in the computation of the surface integrals. 
The details of this procedure were explained in Paper III.

\section{Structure of the 3.5 PN equations of motion}
\label{sec:structure}

In the following sections, we shall 
derive an acceleration for two spherical compact stars 
through the third and a half post-Newtonian accuracy. 
For this purpose, we evaluate the surface integrals in Eqs. 
(\ref{EvolOfFourMom}) and (\ref{generaleom}) to the appropriate order. 
As the mass of the star is $O(\epsilon^2)$, 
the Newtonian force appears as $\epsilon^2$ correction to the lowest
order equations of motion ($m_A d v_A^i/d\tau$ = 0). The 3.5 PN order correction, or
$(v/c)^7$ correction to the Newtonian force appears 
at $O(\epsilon^7)$ and hence the equations that we have to evaluate to derive
an evolution equation for the energy and the   
equations of motion are 
\begin{eqnarray}  
\left(\frac{dP_{1\Theta}^{\tau}}{d\tau}\right)_{\le 3.5 {\rm PN}} &=& 
\left(\frac{dP_{1\Theta}^{\tau}}{d\tau}\right)_{\le 3 {\rm PN}} + 
\epsilon^7 \left[- 
\oint_{\pa B_1}dS_k\,\,\mbox{}_{11}\Theta_{NZ}^{\tau k} 
+ v_1^k 
\oint_{\pa B_1}dS_k\,\,\mbox{}_{11}\Theta_{NZ}^{\tau \tau}
\right], 
\label{EvolOfFourMom3PN}
\\ 
m_1 \left(\frac{d v_1^i}{d\tau}\right)_{\le 3.5 {\rm PN}} 
&=& m_1\left(\frac{dv_1^i}{d\tau}\right)_{\le 3 PN} + 
\epsilon^7 \left[- 
\oint_{\pa B_1}dS_k\,\,\mbox{}_{11}\Theta_{NZ}^{ki} 
+ v_1^k 
\oint_{\pa B_1}dS_k\,\,\mbox{}_{11}\Theta_{NZ}^{\tau i}
\right] 
\nonumber \\
\mbox{} &&
+ \epsilon^7 
\left(\frac{dP_{1\Theta}^{\tau}}{d\tau}\right)_{3.5 {\rm PN}} v_1^i
+ \epsilon^7 \left(
(m_1 - P_{1\Theta}^{\tau})\frac{d v_1^i}{d\tau}\right)_{3.5 {\rm PN}} 
\nonumber \\
\mbox{} &&
- \epsilon^7 \frac{d\, \mbox{}_7 Q_{1\Theta}^i}{d\tau} 
- \epsilon^7 \frac{d^2\, \mbox{}_5 D_{1\Theta}^i}{d\tau^2},     
\label{generaleom3PN}
\end{eqnarray}
where for an equation or a quantity $f$, 
$(f)_{\le n PN}$ 
and $(f)_{n PN}$ denote   
$f$ up to the $n$ PN order inclusively and $f$ at the $n$ PN order, 
respectively. 
$\mbox{}_{\le n}f$ and  
$\mbox{}_n f$ on the other hand denote an equation or a 
quantity $f$ up to $O(\epsilon^n)$ and at $O(\epsilon^n)$, respectively. 
In Paper II, we found $Q_{A\Theta}^i = O(\epsilon^6)$. 
It should be understood that 
in the second line of Eq. (\ref{generaleom3PN}), the acceleration 
$dv_1^i/d\tau$ should be replaced by the acceleration of 
an appropriate order lower than the 3.5 PN order.  
Henceforth, we call Eq. (\ref{generaleom3PN}) the general form 
of the 3.5 PN equations of motion.

The explicit forms of the integrands 
$\mbox{}_{11}\Theta_{NZ}^{\mu\nu} = 
\mbox{}_{11}[-gt_{LL}^{\mu\nu}]$ (on $\pa B_A$) 
are 
\begin{eqnarray}
16\pi\,\mbox{}_{11}\Theta_{NZ}^{\tau\tau} &=&     
-\frac{7}{4} {\mbox{}_4h^{\tau\tau}}_{,k} \,\,{\mbox{}_9h^{\tau\tau}}^{,k}
+\cdots, 
\\
16\pi\,\mbox{}_{11}\Theta_{NZ}^{\tau i} &=&     
2 {\mbox{}_4h^{\tau\tau}}_{,k}\,\,
   {\mbox{}_9h^{\tau}}^{[k,i]} + 
\cdots, \\
16\pi\,\mbox{}_{11}\Theta_{NZ}^{ij} &=&     
\frac{1}{4}
(\delta^i\mbox{}_k\delta^j\mbox{}_l + \delta^i\mbox{}_l\delta^j\mbox{}_k - 
\delta^{ij}\delta_{kl})
\left\{
\mbox{}_4h^{\tau\tau,k}
(\mbox{}_{11}h^{\tau\tau,l}
+ \mbox{}_9h^m\mbox{}_m\mbox{}^{,l} 
+ 4\mbox{}_9h^{\tau l} \mbox{}_{,\tau}
) +
8 \mbox{}_4h^{\tau}\mbox{}_m\mbox{}^{,k}\,\,\mbox{}_9h^{\tau [l,m]}
\right\}
\nonumber \\
\mbox{} &+&  2\mbox{}_4h^{\tau i}\mbox{}_{,k}\,\,\mbox{}_9h^{\tau [k,j]}
+ 2\mbox{}_4h^{\tau j}\mbox{}_{,k}\,\,\mbox{}_9h^{\tau [k,i]}
+ \cdots,  
\label{eq:tLLIJ11}
\end{eqnarray}
See the appendix \ref{sec:appendix_tLL} for the complete expressions.
The field components up to the 2.5 PN order inclusively, 
$\mbox{}_{\le 9}h^{\tau\tau}$, $\mbox{}_{\le 7}h^{\tau i}$, 
and $\mbox{}_{\le 7}h^{ij}$, are listed in Paper II.
In addition to those, to derive the 3.5 PN mass-energy relation and 
the 3.5 PN momentum-velocity relation, we have to derive 
$\mbox{}_9h^{\tau i}$. To derive the 3 PN equations of motion, 
we further need $\mbox{}_{11}h^{\tau\tau} + \mbox{}_9h^k\mbox{}_k$.   
We do not need to use the 3 PN field 
$\mbox{}_{8}h^{\tau i}$, $\mbox{}_{8}h^{ij}$, 
and $\mbox{}_{10}h^{\tau \tau}$ for the purpose of this paper.

Up to the 2.5 PN order inclusively, the super-potentials required  
to compute the field could be found \cite{BFP98,JS98}. 
At the 3 PN order, it is quite 
difficult to complete the required super-potentials. We took  
another method to overcome this problem in Paper III. 
Fortunately, we can find the super-potentials at the 3.5 PN order and hence 
derive the 3.5 PN field in a closed form. 
The next section shows our method of the derivation of the 3.5 PN field.

\section{3.5 PN Gravitational Field in Harmonic Coordinates}
\label{sec:field}

As written in the previous section, our derivation of the 3.5 PN equations of motion 
requires $\mbox{}_9h^{\tau i}$ and 
$\mbox{}_{11}h^{\tau\tau} + \mbox{}_9h^k\mbox{}_k$.  We start with the
body zone contribution.

\subsection{3.5 PN Body Zone Contribution for a Spherically Symmetric Star}

This paper studies equations of motion for a spherically symmetric star
and hence discard all the GFC multipole moments of the stars. 
As stated 
in Sec. \ref{sec:sphericalStar} and will be shown in Sec. \ref{LorentzContraction}, 
up to the 3.5 PN order we can safely discard the NZC multipole moments
$I_A^{K_l}$, $J_A^{(K_{l-1}[k_l)i]}$ 
and $Z_A^{(K_{l-1}[k_l)i]j}$ except 
for the NZC
quadrupole moments $I_A^{ij}$. Using Eqs. (\ref{MomVelRelation}) to
(\ref{ZLijToJLij}), 
the body zone contribution for a spherically symmetric star becomes  
\begin{eqnarray}
h^{\tau\tau}_B &=& 4 \epsilon^4 \sum_{A=1,2}
\left[
\frac{P_A^{\tau}}{r_A} + \epsilon^2 \frac{D_A^k r^k_A}{r_A^3} +
 \epsilon^4 \frac{3 I_A^{kl} r^{<kl>}_A }{2 r_A^5} 
\right] 
\nonumber \\
&
+& 4 \sum_{n=2}^{7} \frac{(-\epsilon)^{4 + n}}{n!} 
\frac{\partial^n}{\partial\tau^n} 
\left[
\sum_{A=1,2}P_A^{\tau}r_A^{n-1}
\right]
 + O(\epsilon^{12}), 
\label{hBttInAp} 
\end{eqnarray} 
\begin{eqnarray}
h^{\tau i}_B &=& 4 \epsilon^4 \sum_{A=1,2}
\left[
\frac{P_A^{\tau}v_A^i }{r_A}
+ \epsilon^2 \frac{1}{r_A}\frac{dD_A^{i}}{d\tau} 
+ \sum_{n=4}^5\epsilon^n \frac{\mbox{}_nQ_A^i}{r_A}
\right]
\nonumber \\
&
+& 4 \sum_{n=2}^{7} \frac{(-\epsilon)^{4 + n}}{n!} 
\frac{\partial^n}{\partial\tau^n} 
\left[
\sum_{A=1,2}P_A^{\tau}v_A^i r_A^{n-1}
\right]
 + O(\epsilon^{10}),
\label{hBtiInAp} 
\end{eqnarray} 
\begin{eqnarray}
h^{ij}_B &=& 4 \epsilon^4 \sum_{A=1,2}
\left[
\frac{P_A^{\tau}v_A^{i}v_A^j}{r_A}
+ \sum_{n=4}^5 \epsilon^{n}\frac{\mbox{}_nQ_A^{(i}v_A^{j)}}{r_A}
\right.
\nonumber \\
\mbox{} &&+ \left.
\sum_{n=4}^5 \epsilon^{n}
\left(
\frac{5\mbox{}_nR_A^{klm(ij)}r_A^{<klm>}}{8r_A^7}
+ \frac{\mbox{}_nR_A^{(ij)}}{r_A}
+ \frac{r^k_A}{2 r_A^3}(\mbox{}_nR_A^{kji} + 
\mbox{}_nR_A^{kij} - 
\mbox{}_nR_A^{ijk} ) 
 \right)
 \right]
\nonumber \\
\mbox{} &-& 
4 \epsilon^5 \frac{\pa}{\pa \tau}
\sum_{A=1,2}\left[
P_A^{\tau}v_A^iv_A^j 
+ 
\epsilon^4\left(
\mbox{}_4Q_A^{(i}v_A^{j)} + \mbox{}_4R_A^{(ij)} +
\frac{1}{2}\frac{d \mbox{}_4Q_A^{ij}}{d\tau}
\right)
\right]
\nonumber \\
&+&
 4 \sum_{n=2}^{7} \frac{(-\epsilon)^{4 + n}}{n!} 
\frac{\partial^n}{\partial\tau^n} 
\left[
\sum_{A=1,2}P_A^{\tau}v_A^i v_A^j r_A^{n-1}
\right]
+  O(\epsilon^{10}).
 \label{hBijInAp}
\end{eqnarray}
In the above equations, $D_A^i$ may be specified freely to define 
the star's representative point $z_A^i$. 
Although we explicitly wrote $Q_A^{K_li}$ 
and $R_A^{K_lij}$ integrals above, the next section shall show those do not affect the 
3.5 PN acceleration.

\subsubsection{$Q_A^{K_li}$ and $R_A^{K_lij}$}
\label{sec:QandRIntegral}

The surface integrals $Q_A^{K_li}$ and $R_A^{K_lij}$ 
did contribute to the field and the equations of motion 
at the 3 PN order (Paper III). 

At the 3.5 PN order, we compute 
$Q_A^{K_li}$ and $R_A^{K_lij}$ up to $O(\epsilon^5)$ to
derive the 3.5 PN field. Then the integrands of the surface integrals in
Eqs. (\ref{QL}) and (\ref{RL}) 
are  $\mbox{}_9\Lambda_{NZ}^{\mu\nu}$ . Now suppose those integrands behaves 
near the star $A$ as $1/r_A^p$ with $p$ some positive integer. Then 
$Q_A^{K_li} = O((\epsilon R_A)^{l+3-p})$ and $R_A^{K_lij} = O((\epsilon
R_A)^{l+4-p})$. Because we neglect any explicit terms that depend on
$(\epsilon R_A)^q$ ($q$: non-zero integer. See Sec. \ref{Arbitrariness} and Sec. III E of Paper III), we
compute only $Q_A^{K_li}$ for $0\le l \le p-3$ and $R_A^{K_lij}$ for $0 \le l \le
p-4$. Inspection shows that $p=6$ for $\mbox{}_9\Lambda_{NZ}^{\tau\mu}$ and 
$p=7$ for $\mbox{}_9\Lambda_{NZ}^{i \mu}$. For those degree $l$ of the collective
indexes $K_l$,  by explicitly computing those 
integrals in Eqs. (\ref{QL}) and (\ref{RL}),  
we found that $Q_A^{K_li}$ and $R_A^{K_lij}$ 
do not contribute to the field nor equations of motion at the 3.5 PN order.

\subsubsection{A Spherical Body and Multipole Moments}
\label{LorentzContraction}

As promised, this section explains the reason 
we can safely discard the NZC multipole moments except for the NZC
quadrupole moments through the 3.5 PN order. Denoting the NZC (GZC) quadrupole
moment as $I_{A,{\rm NZC}}^{ij}$ ($I_{A,{\rm GFC}}^{ij}$), 
Eq. (C14) of Paper III 
gave an expression of the difference between these two
moments;
\begin{eqnarray}
\delta I_A^{ij} &\equiv& I_{A,{\rm NZC}}^{ij} - I_{A,{\rm GFC}}^{ij} 
=  
\epsilon^{-8}\left(B^i(\tau)v_A^j - A^{ij}(\tau)\right)
\oint_{\partial B_A}dS_{m}\,y_A^iy_A^jy_A^m\Lambda_{NZ}^{\tau\tau}, 
\label{eq:diffNZC_GFC}
\end{eqnarray}
for a star which is spherically symmetric in the GFC.  
The coefficients $A^{ij}(\tau)$ and $B^i(\tau)$ can be read off from \cite{AB86} 
(see also Eq. (C16)
and (C17) of Paper III) and are 
\begin{eqnarray}
B^i(\tau)v_1^j &=& \epsilon^2 v_1^iv_1^j + O(\epsilon^4),\\
A^{ij}(\tau) &=& \epsilon^2 \left(\frac{1}{2}v_1^iv_1^j -
			     \frac{m_2}{r_{12}}\delta^{ij}\right) +
O(\epsilon^4),  
\end{eqnarray} 
for the star $A=1$ and $\vec r_{12} = \vec z_1 - \vec z_2 = \vec r_2 -
\vec r_1$. Note that there is no $\epsilon^3$ term in
$A^{ij}(\tau)$ nor $B^i(\tau)$ because there is no 0.5 PN field in the
harmonic gauge (i.e., $\mbox{}_5h^{\tau\tau} = 0$)  as was shown in the appendix E of Paper II.
The surface integral in Eq. (\ref{eq:diffNZC_GFC}) can be expanded in
$\epsilon$ as 
\begin{eqnarray}
\epsilon^{-8}\oint_{\partial B_A}dS_{m}\,y_A^iy_A^jy_A^m\Lambda_{NZ}^{\tau\tau} &=& 
\oint_{\partial B_A}dS_{m}\,y_A^iy_A^jy_A^m\mbox{}_8\Lambda_{NZ}^{\tau\tau}
+ \epsilon
\oint_{\partial B_A}dS_{m}\,y_A^iy_A^jy_A^m\mbox{}_9\Lambda_{NZ}^{\tau\tau}
\nonumber \\
&+& O(\epsilon^2).
\end{eqnarray} 
The first integral in the right hand side of the equality is found to be non-zero and contribute to the 3 PN
metric $\mbox{}_{10}h^{\tau\tau}$ and hence the 3 PN equations of
motion. The second integral could in principle affect the 3.5 PN field, as
the coefficient $\epsilon$ in front of it indicates. So let us evaluate the
second integral. Its integrand is 
$\mbox{}_9\Lambda_{NZ}^{\tau\tau} = -
\mbox{}_4h^{\tau\tau}\mbox{}_{,ij}\mbox{}_5h^{ij}$ which behaves near
the star $A=1$ as $\mbox{}_9\Lambda_{NZ}^{\tau\tau} \sim 3 r_1^i
r^j_1/r_1^5 - \delta^{ij}/r_1^3$. The surface integral then results in 
\begin{eqnarray}
\oint_{\partial B_A}dS_{m}\,y_A^iy_A^jy_A^m\mbox{}_9\Lambda_{NZ}^{\tau\tau} 
&=& O(\epsilon^2 R_A^2),
\end{eqnarray} 
and thus there is no monopole term in $\delta I_A^{ij}$ at the 3.5 PN
order. Similarly, using Eq. (C14) of Paper III, 
we can check that none of the NZC multipole moments hide any monopole 
terms at the 3.5 PN order for a star spherically symmetric in the GFC.

\subsection{3.5 PN N/B Field}
\label{sec:3hPNNBField}

This section explains how we derive the 3.5 PN $N/B$ field components 
$\mbox{}_9h_{NZ/B}^{\tau i}$ 
and 
$\mbox{}_{11}h_{NZ/B}^{\tau\tau} + \mbox{}_9h_{NZ/B}^k\mbox{}_k$. 
The integrals for the latter which we evaluate may be written as   
\begin{eqnarray}
\lefteqn{
\mbox{}_{11} h_{NZ/B}^{\tau \tau}(\tau,\vec x) + \mbox{}_{9}
h_{NZ/B}^k\mbox{}_k(\tau,\vec x)} \nonumber \\
 &=&  
4\sum_{n=0,n\ne 1}^{5} \frac{(-1)^n}{n!}\frac{\pa^n}{\pa \tau^n}
\int_{NZ/B}  d^3y 
\frac{
\mbox{}_{11 - n}\Lambda_{NZ}^{\tau\tau}(\tau,\vec{y}) + 
\mbox{}_{9-n}\Lambda_{NZ}^k\mbox{}_k(\tau,\vec{y}) 
}
{|\vec x-\vec y|^{1-n}}
\nonumber \\
&-&
 4 
\int_{\partial NZ}  dS_j\, 
\mbox{}_{10}\Lambda_{NZ}^{\tau j}(\tau,\vec{y})  
- 4 \frac{\pa}{\pa \tau}
\int_{NZ/B}  d^3y\,
\mbox{}_{8}\Lambda_{NZ}^k\mbox{}_k(\tau,\vec{y}). 
\label{eq:h11tt9kk}
 \end{eqnarray}
Because only a spatial derivative of 
$\mbox{}_{11}h_{NZ/B}^{\tau\tau} + \mbox{}_9h_{NZ/B}^k\mbox{}_k$
appears in $\mbox{}_{11}[-gt_{LL}^{ij}]$ as shown in
Eq. (\ref{eq:tLLIJ11}),  
the second and the third term in the right hand of the equality in the above equation does
not contribute to the 3.5 PN equations of motion. Similarly,
$\mbox{}_9h_{NZ/B}^{\tau i}$ may be written as 
\begin{eqnarray}
\mbox{}_{9} h_{NZ/B}^{\tau i}(\tau,\vec x)
 &=&  
4 \int_{NZ/B}  d^3y 
\frac{
\mbox{}_9\Lambda_{NZ}^{\tau i}(\tau,\vec{y})  
}
{|\vec x-\vec y|}
-4 \int_{\partial NZ}  dS_j\,
\mbox{}_8\Lambda_{NZ}^{ij}(\tau,\vec{y})  
\nonumber \\
&-&
\frac{2}{3} \frac{\pa^3}{\pa \tau^3}
\int_{NZ/B}  d^3y\, 
\mbox{}_6 \Lambda_{NZ}^{\tau i}(\tau,\vec{y})  
|\vec x-\vec y|^3.
\label{eq:h9ti}
 \end{eqnarray}
Our task now is to find super-potentials $f$ that satisfies, e.g.,    
$\mbox{}_{11}\Lambda_{NZ}^{\tau\tau}+ 
\mbox{}_{9}\Lambda_{NZ}^k\mbox{}_k = \Delta f$. 
With the super-potentials, 
we use Eq. (\ref{NBcontribution}) and find explicit 
expressions of the 3.5 PN field components in closed forms.  
For the higher order retarded expansion terms, for example for 
the second retarded expansion term of Eq. (\ref{eq:h11tt9kk})
(the $n=2$ term of the first term in Eq. (\ref{eq:h11tt9kk}))
, we use 
super-super-potential $f(\vec y)$ satisfying $\mbox{}_9\Lambda_{NZ}^{\tau\tau} + 
\mbox{}_7\Lambda_{NZ}^{k}\mbox{}_k  = \Delta^2 f$.  Then the volume
integral may be converted into the bulk term and the surface integral
terms as  
\begin{eqnarray*}
\lefteqn{\int_{NZ/B}d^3y|\vec x - \vec y| \Delta^2 f(\vec y)}
\nonumber \\
&=& - 8\pi f(\vec x) 
\nonumber \\
\mbox{} &&+ 
\oint_{\pa (NZ/B)}dS_k\left[
|\vec x - \vec y|\pa_k\Delta f(\vec y) 
- \frac{y^k-x^k}{|\vec x - \vec y|}\Delta f(\vec y) 
+ \frac{2}{|\vec x - \vec y|}\pa_k f(\vec y)  
+ \frac{2(y^k-x^k)}{|\vec x - \vec y|^3} f(\vec y) 
\right].  
\end{eqnarray*}

The source terms for which we need to find particular solutions of 
Poisson equations have the following spatial coordinate dependence. 
\begin{eqnarray}
&&
\left\{\frac{1}{r_1^6},
\frac{1}{r_1^5},
\frac{1}{r_1^4},
\frac{1}{r_1^3},
\frac{1}{r_1^2},
\frac{1}{r_1},
r_1,
\right. \nonumber \\&& \left. 
\frac{{r_1^i}}{r_1^6},
\frac{{r_1^i}}{r_1^5},
\frac{{r_1^i}}{r_1^4},
\frac{{r_1^i}}{r_1^3},
\frac{{r_1^i}}{r_1},
\frac{{r_1^i}{r_1^j}}{r_1^7},
\frac{{r_1^i}{r_1^j}}{r_1^5},
\frac{{r_1^i}{r_1^j}}{r_1^3},
\frac{{r_1^i}{r_1^j}{r_1^k}}{r_1^7},
\frac{{r_1^i}{r_1^j}{r_1^k}}{r_1^5},
\right. \nonumber \\&& \left. 
\frac{r_1^2}{{r_2}^6},
\frac{r_1^4}{{r_2}^6},
\frac{r_1}{{r_2}^5},
\frac{r_1^2}{{r_2}^5},
\frac{r_1^3}{{r_2}^5},
\frac{r_1^4}{{r_2}^5},
\frac{r_1^6}{{r_2}^5},
\frac{r_1^2}{{r_2}^4},
\frac{r_1}{{r_2}^3},
\frac{r_1^2}{{r_2}^3},
\frac{r_1^4}{{r_2}^3},
\frac{r_1^2}{{r_2}},
\frac{1}{r_1^5 {r_2}^3},
\frac{1}{r_1^5 {r_2}},
\frac{1}{r_1^3{r_2}},
\right. \nonumber \\&& \left. 
\frac{{r_1^i}}{{r_2}^5},
\frac{r_1^2 {r_1^i}}{{r_2}^5},
\frac{r_1^4 {r_1^i}}{{r_2}^5},
\frac{{r_1^i}}{{r_2}^3},
\frac{{r_1^i}}{r_1^5 {r_2}^3},
\frac{{r_1^i}}{r_1^3 {r_2}^3},
\frac{{r_1^i}}{r_1 {r_2}^3},
\frac{r_1^2 {r_1^i}}{{r_2}^3},
\frac{{r_1^i}}{{r_2}},
\frac{{r_1^i}}{r_1^3 {r_2}},
\frac{{r_1^i} {r_2}}{r_1^5},
\frac{{r_1^i} {r_2}^2}{r_1^6},
\frac{{r_1^i} {r_2}^2}{r_1^5},
\frac{{r_1^i} {r_2}^2}{r_1^3},
\right. \nonumber \\&& \left. 
\frac{{r_1^i}{r_1^j} {r_2}^2}{r_1^7},
\frac{{r_1^i}{r_1^j} {r_2}^2}{r_1^5},
\frac{{r_1^i}{r_1^j}{r_1^k} {r_2}^2}{r_1^7},
\frac{{r_1^i} {r_2}^4}{r_1^5},
\frac{{r_1^i}{r_1^j} {r_2}^4}{r_1^7},
\frac{{r_1^i}{r_2^i}}{r_1^5},
\frac{{r_1^i} {r_2^i}}{r_1^3},
\frac{{r_1^i}{r_1^j} {r_2^k}}{r_1^5},
\frac{r_1^2 {r_1^i}{r_2^j}}{{r_2}^5},
\frac{1}{r_1^3 {r_2}^3}\right\}, 
\end{eqnarray}
and $1 \leftrightarrow 2$, i.e., like $1/r_2^2$ 
for $1/r_1^2$. (Specific members of the list depends on how one
simplifies the expressions of the source terms of the Einstein equations. For instance, 
one can always erase $r_2^i$ by using $r_2^i = r_{12}^i + r_1^i$.)

Our method to derive the super-potentials are heuristic;  
there are few guidelines available to find the required 
super-potentials. We proceed as follows.
First, we convert all the tensorial sources into scalars   
with spatial derivatives. For example, $\mbox{}_{11}\Lambda_{NZ}^{\tau\tau}+ 
\mbox{}_{9}\Lambda_{NZ}^k\mbox{}_k$ include the following term  
\begin{eqnarray}
\frac{288 m_1 m_2^2 r_1^2}{{r_2}^5}(\vec r_1\cdot\vec v_1)(\vec r_2\cdot\vec V)(\vec r_{12}\cdot\vec V)
&=& 
- 24 m_1 m_2^2 v_1^i V^j (\vec r_{12}\cdot\vec V)
\frac{\partial }{\partial z_1^i}\frac{\partial }{\partial z_2^j}
\left(\frac{r_1^4}{r_2^3}\right)
\label{eq:superpotentialexample1}.  
\end{eqnarray}
(Here and henceforth, it should be understood that 
in general 
``scalars'' can have tensorial indexes carried 
by $\vec v_A$ and $\vec r_{12}$, but 
do not have those by $\vec r_A$.)

Second, we find the particular solutions for Poisson 
equations with the scalars as sources using a formula 
$\Delta (f(\vec x)g(\vec x)) 
= g(\vec x)\Delta f(\vec x) + 
2 \vec \nabla f(\vec x) \cdot \vec \nabla g(\vec x) 
+ f(\vec x)\Delta g(\vec x)$ 
valid in $NZ/B$.  
We also use   
{\it super-potential chains} such as; 
\begin{center}
\begin{tabular}{ccc}
$f^{(-3,-2)}$ & 
$\stackrel{\Delta_{11}}{\longrightarrow}$ & 
$6 f^{(-5,-2)}$\\
$\downarrow {\scriptstyle \Delta_{22}}$ & 
  & 
$\downarrow {\scriptstyle \Delta_{22}}$\\ 
$2 f^{(-3,-4)}$ &
$\stackrel{\Delta_{11}}{\longrightarrow}$ & 
$12 f^{(-5,-4)}$,  
\end{tabular}
\end{center}
where 
$$f^{(-3,-2)} =\frac{1}{r_1r_{12}^2}\ln\left(\frac{r_2}{r_1}\right),$$
and $f^{(m,n)}$ satisfies $\Delta f^{(m,n)} = r_1^m r_2^n$. 
$\Delta_{AA'} = \partial^2/\partial z_A^i/\partial z_{A'i}$. 
For example, 
for Eq. (\ref{eq:superpotentialexample1}), 
it is easy to find a particular solution, 
\begin{eqnarray*}
\frac{r_1^2 r_1^i r_2^j}{r_2^5} &=& 
\Delta\left[
-\frac{1}{12}\frac{\partial }{\partial z_1^i}\frac{\partial }{\partial z_2^j}
f^{(4,-3)} \right],  
\end{eqnarray*}
where 
\begin{eqnarray}
f^{(4,-3)} &=& - \frac{r_1^4}{4r_2} - \frac{r_1^2r_{12}^2}{2r_2} +
 r_1^2r_2 + 2r_{12}^2r_2 - \frac{2 r_2^3}{3} -\frac{r_{12}^4}{r_2}\ln
 r_2. 
\end{eqnarray}

Following the method described above, 
we could find all the required particular solutions and 
using algebraic computation codes written in Mathematica    
we have derived the expressions 
of $\mbox{}_9h^{\tau i}$ and $\mbox{}_{11}h^{\tau \tau} + \mbox{}_9h^k\mbox{}_k$ 
in closed forms. However  
we do not reproduce those in this paper because  
the numbers of terms in the expressions   
are huge ($\sim 800$ and $\sim 1000$,
respectively. These numbers depend on 
a particular simplification one makes on those expressions).

Incidentally, we have checked (a part of) the 3.5 PN harmonic condition 
$\mbox{}_{\le 9}h^{\nu \mu}\mbox{}_{,\mu} =0$.

\section{3.5 PN Mass-Energy Relation in Harmonic Coordinates}
\label{sec:mass-energy}

By evaluating 
the surface integrals in the evolution equation of the star's energy Eq. (\ref{EvolOfFourMom3PN}), 
one may obtain the time derivative of the mass of the star 1 as 
\begin{eqnarray}
\lefteqn{
\left(\frac{d P_{1 \Theta}^{\tau}}{d \tau}\right)_{\le 3.5PN}
= 
\left(\frac{d P_{1 \Theta}^{\tau}}{d \tau}\right)_{\le 3PN}
} 
\nonumber \\
&+&  
\frac{8{m_1}^4 {m_2} }{15 {r_{12}}^5}
+\frac{56 {m_1}^3 {m_2}^2 }{15 {r_{12}}^5}
+\frac{16 {m_1}^2{m_2}^3 }{5 {r_{12}}^5}
\nonumber \\
&-&
\frac{m_1^2m_2^2}{r_{12}^4}\left[
\frac{148}{15}{v_1}^2 
+\frac{392}{15} {(\vec n_{12}\cdot\vec v_1)}^2 
+\frac{184}{15} {(\vec n_{12}\cdot\vec v_2)}^2
-\frac{52}{15} 
   {v_2}^2 
\right. \nonumber \\&& \left.  
-\frac{192}{5}  {(\vec n_{12}\cdot\vec v_1)} {(\vec n_{12}\cdot\vec v_2)} 
+\frac{40}{3}  {(\vec v_{1}\cdot\vec v_2)} 
\right]
\nonumber \\
&+&
\frac{m_1^3m_2}{r_{12}^4}
\left[
-\frac{16}{5}{(\vec n_{12}\cdot\vec v_1)}^2 
+\frac{8}{5} {(\vec n_{12}\cdot\vec v_2)}^2 
+\frac{4}{5} {v_1}^2 
- \frac{4}{5} {v_2}^2 
+\frac{8}5  {(\vec n_{12}\cdot\vec v_1)}
   {(\vec n_{12}\cdot\vec v_2)}\right]
\nonumber \\ &+&
\frac{m_1^2m_2}{r_{12}^3}\left[
-\frac{8}{15}  {v_1}^4 
+\frac{4}{15}  {v_2}^4 
+\frac{8}{5}  {(\vec n_{12}\cdot\vec v_1)}^2 {v_1}^2 
-\frac{4}{5}  {(\vec n_{12}\cdot\vec v_2)}^2 {v_1}^2 
-\frac{4}{5}  {(\vec n_{12}\cdot\vec v_1)} {(\vec n_{12}\cdot\vec v_2)} {v_1}^2
\right. 
\nonumber \\
&&-\left. 
\frac{8}{15}  {(\vec v_{1}\cdot\vec v_2)}^2 
+\frac{8}{5} {(\vec n_{12}\cdot\vec v_1)}^2 {v_2}^2 
-\frac{4}{5} {(\vec n_{12}\cdot\vec v_2)}^2 {v_2}^2 
-
\frac{4}{15} {v_1}^2 {v_2}^2 
-\frac{4}{5} {(\vec n_{12}\cdot\vec v_1)} {(\vec n_{12}\cdot\vec v_2)} {v_2}^2 
\right.
\nonumber \\ &&\left.
-\frac{4}{15} {(\vec v_{1}\cdot\vec v_2)} {v_2}^2 
-\frac{16}{5}{(\vec n_{12}\cdot\vec v_1)}^2
   {(\vec v_{1}\cdot\vec v_2)} 
+ \frac{8}{5} {(\vec n_{12}\cdot\vec v_2)}^2 {(\vec v_{1}\cdot\vec v_2)} 
+\frac{4}{3} {v_1}^2 {(\vec v_{1}\cdot\vec v_2)}
\right. 
\nonumber \\ && \left. 
+\frac{8}{5} {(\vec n_{12}\cdot\vec v_1)} {(\vec n_{12}\cdot\vec v_2)} {(\vec v_{1}\cdot\vec v_2)} 
\right].
\label{THPNEMSurfaceIntegralFormTimeResult}
\end{eqnarray}
Paper III gives the explicit expression of 
$\left({d P_{1 \Theta}^{\tau}}/{d \tau}\right)_{\le 3PN}$.

We can integrate Eq. 
(\ref{THPNEMSurfaceIntegralFormTimeResult}) functionally as 
\begin{eqnarray}
P^{\tau}_{1 \Theta} &=& m_1 
\sum_{k = 0}^{7} \epsilon^k \mbox{}_k\Gamma_1 + 
O(\epsilon^8).
\label{3hPNEnergy}
\end{eqnarray}
The $\mbox{}_k\Gamma_A$ up to 3 PN order are given in Paper III. 
The new result we show in this paper is $\mbox{}_7\Gamma_A$; 
\begin{eqnarray}
\mbox{}_7\Gamma_1  &=& 
-\frac{8 {m_1}^2{m_2} {(\vec n_{12}\cdot\vec V)} }{15 {r_{12}}^3}
+\frac{4 {m_1}{m_2} {(\vec n_{12}\cdot\vec V)} V^2 }{15 {r_{12}}^2}
-\frac{16 {m_1}{m_2}^2 {(\vec n_{12}\cdot\vec V)} }{5 {r_{12}}^3}, 
\end{eqnarray}
where $\vec V = \vec v_1 - \vec v_2$. 
The mass-energy relation for the $\chi$ part up to the 3.5 PN 
order is given in the appendix \ref{sec:chipart}.
Eqs. (\ref{3hPNEnergy}) and (\ref{Ptchi3hPN}) 
give the 3.5 PN order mass-energy relation in our formalism. 

From the definition of $P_{A\Theta}^{\tau}$, 
we expect it equal to $\sqrt{-g} m_A u^{\tau}_A$ where 
$u^{\tau}_A$ is the time component of the 4-velocity of 
the star $A$ normalized as 
$g_{\mu\nu}u_A^{\mu}u_A^{\nu} = -\epsilon^{-2}$ with 
$u_A^i = u_A^{\tau}v_A^i$. 
$\sqrt{-g} u^{\tau}_A$ at 3.5 PN order can be written in terms of the
deviation field as 
\begin{eqnarray}
\sqrt{-g} u_{A}^{\tau} 
&=& 
\mbox{}_{\le 6}(\sqrt{-g} u_{A}^{\tau}) \nonumber \\
&+&
\epsilon^7 
\left(\frac{3}{4}\mbox{}_9h^{\tau\tau} - 
\frac{1}{4}\mbox{}_7h^k\mbox{}_k
- \mbox{}_7h^{\tau i}v_A^i
+\frac{7}{8}\mbox{}_7h^{\tau\tau}v_A^2
-\frac{3}{16}\mbox{}_4h^{\tau\tau}\mbox{}_7h^{\tau\tau}
\right. \nonumber \\ &&\left.
+ \frac{1}{2}\mbox{}_5h_{ij}v_A^iv_A^j 
- \frac{1}{2}\mbox{}_5h^k\mbox{}_kv_A^2
+ \frac{1}{16} \mbox{}_5h^k\mbox{}_k\mbox{}_4h^{\tau\tau}
\right)
+ O(\epsilon^8).
\end{eqnarray} 
This expression should be somehow evaluated at the star $A$. 
However, because the deviation field  $h^{\mu\nu}$ diverges at the star in the 
point particle description, the above expectation is not trivial. 
As in Paper I, II, and III,  
we checked that up to the 3.5 PN order inclusively  a relation 
\begin{eqnarray}
P_{A \Theta}^{\tau} = m_A[\sqrt{-g}u^{\tau}_A]_{A}^{ext},  
\label{MassEnergyRelationFull}
\end{eqnarray}
holds where $[f]^{ext}_{A}$
means that we regularize  
the quantity f at the star $A$  
by the Hadamard's Partie Finie (see e.g. \cite{BF00}) 
or whatever regularization which 
gives the same result. We here emphasize that 
we have never assumed this ``natural''  
relation in advance. The relation Eq. 
(\ref{MassEnergyRelationFull}) has been
derived by solving the evolution equation for
$P_{A \Theta}^{\tau}$ functionally. 

For the 3.5 PN mass-energy relation Eqs. (\ref{3hPNEnergy}), 
one can check that only the field components up to the 2.5 PN order 
$\mbox{}_{\le 9}h^{\tau\tau}$ and $\mbox{}_{\le 7}h^{i\nu}$ 
appear in the expression of $\sqrt{-g}u^{\tau}_A$ 
in Eq. (\ref{MassEnergyRelationFull}) above.
We apply a regularization on those components in the right hand side of 
Eq. (\ref{MassEnergyRelationFull}). 
The ``naturality'' of 
Eq. (\ref{MassEnergyRelationFull}) thus supports a use of the 
Hadamard Partie Finie regularization
in the literature up to the 2.5 PN order \cite{BFP98}.

\section{3.5 PN Momentum-Velocity Relation in Harmonic Coordinates}
\label{sec:momentum-velocity}

From Eq. (\ref{MomVelRelation}), an explicit expression of 
the momentum-velocity relation is obtained by evaluating 
the $Q_{A\Theta}^i$ integral up to $O(\epsilon^7)$ 
\begin{eqnarray} 
Q_{A\Theta}^i &=&   
\mbox{}_{\le 6} Q_{A\Theta}^i  + 
\epsilon^{7}
 \oint_{\pa B_A} dS_k
 \left(\mbox{}_{11}[-g t_{LL}^{\tau k}]  - 
\mbox{}_{11}[-g t_{LL}^{\tau\tau}] v_A^k  \right) y_A^i. 
\end{eqnarray} 
The computation is straightforward and we found 
$\mbox{}_7Q_{A\Theta}^i = 0$. From the 3 PN accurate  
$Q_{A\Theta}^i$ calculation, we have a momentum-velocity relation.   
\begin{eqnarray}
P_{1\Theta}^i 
&=& P_{1\Theta}^{\tau} v_1^i  
- \epsilon^6 \frac{d}{d\tau}\left(\frac{1}{6}m_1^3a_1^i\right) 
+ \epsilon^2 \frac{d D_{1\Theta}^i}{d\tau} + 
O(\epsilon^8),
\label{eq:3hPMMomuntumVelocityRelation}
\end{eqnarray}
where $a_1^i$ is the acceleration of the star 1 and should be replaced
by the 3 PN order expression of the acceleration. 
As in Paper III, we define the representative points of the 
stars $z_A^i$ by choosing 
\begin{eqnarray}
&&
D_{A\Theta}^i(\tau) =  \epsilon^4 \frac{1}{6} m_A^3a_A^i
-   \epsilon^4\frac{22}{3} 
m_A^3 a_A^i \ln \left(\frac{r_{12}}{\epsilon R_A}\right).    
\label{3hPNDefOfCM}
\end{eqnarray}
The first term in Eq. (\ref{3hPNDefOfCM}) 
makes the three momentum proportional to 
$v_A^i$ at the 3 PN order. The second term gauges away the logarithmic terms from 
the 3 PN equations of motion \cite{Itoh:2003fz}. In any case, our choice
of $D_{A\Theta}^i$ affects only the 3 PN order correction and $P_{1\Theta}^i 
= P_{1\Theta}^{\tau} v_1^i$ holds as long as we are concerned with the
3.5 PN order correction to the equations of motion.

As the second term in Eq. (\ref{3hPNDefOfCM}) depends on the arbitrary
parameter $\epsilon R_A$, one may suspect that our equations of
motion lose their predictive power on the binary dynamics.
This is not the case. When 
we define the star's  representative point by $D_{A\Theta}^i = 0$
instead of Eq. (\ref{3hPNDefOfCM}), logarithmic terms that 
depend on $\epsilon R_A$ appear in equations of motion
\cite{Itoh:2003fz}. But those terms are mere gauge. Indeed, 
it can be shown that an observable such as the orbital energy does not depend on $\epsilon R_A$ when that observable is
written in terms of gauge independent variables (such as the
gravitational wave frequency) \cite{Blanchet:2000nv,Itoh:2003fy}.

The shift in the world line induced by the second term in
Eq. (\ref{3hPNDefOfCM}) is 
equivalent to the gauge transformation used in the works by 
Blanchet and Faye \cite{Blanchet:2000nv,Blanchet:2000ub}. 
However, the fact that we adopt Eq. (\ref{3hPNDefOfCM}) does not mean 
that we use a regularization in any sense. The generalized 
Hadamard Partie Finie or the dimensional regularization is nothing to do 
with any mere shift of the world line. In fact, the undetermined
coefficient $\lambda$ associated with their use of the generalized
Hadamard Partie Finie cannot be gauged away by any shift of the world
line \cite{Blanchet:2000nv,Blanchet:2000ub}.

\section{3.5 PN Equations of Motion in Harmonic Coordinates}
\label{sec:eom}

With the 3.5 PN field at hand, we evaluate the surface integrals in the 
3.5 PN general equations of motion Eq. (\ref{generaleom3PN}). A tedious
but straightforward calculation results in 
\begin{eqnarray}
\lefteqn{(m_1 a^i_1)_{\le 3.5{\rm PN}} = (m_1 a^i_1)_{\le 3 {\rm PN}}}
\nonumber \\ &
+& 
\frac{m_1^4m_2}{r_{12}^5}
\left[n_{12}^i \left\{\frac{3992}{105} (\vec n_{12}\cdot\vec v_1)
-\frac{4328}{105}(\vec n_{12}\cdot\vec v_2)\right\}-\frac{184}{21} V^i\right]
 \nonumber \\ &+& 
\frac{m_1^3 m_2^2}{r_{12}^5}
   \left[\frac{6224}{105} V^i+n_{12}^i \left\{\frac{2872}{21} (\vec n_{12}\cdot\vec v_2)-\frac{13576}{105} (\vec n_{12}\cdot\vec v_1)\right\}\right]
\nonumber \\ &+& 
\frac{m_1^3 m_2}{r_{12}^4}
V^i \left[-\frac{132}{35} {v_1}^2-\frac{48}{35} {v_2}^2+\frac{52}{15} (\vec n_{12}\cdot\vec v_1)^2 
+\frac{152}{35} (\vec v_1\cdot\vec v_2)
 \right. \nonumber \\ &&  \left.
-\frac{56}{15} (\vec n_{12}\cdot\vec v_1)(\vec n_{12}\cdot\vec v_2)
-\frac{44}{15} (\vec n_{12}\cdot\vec v_2)^2
\right]
 \nonumber \\ &+&  
\frac{m_1^3 m_2}{r_{12}^4}
n_{12}^i \left[-\frac{4888}{105} (\vec n_{12}\cdot\vec v_1) {v_1}^2
+\frac{5056}{105} (\vec n_{12}\cdot\vec v_2) {v_1}^2-\frac{1028}{21} {v_2}^2
   (\vec n_{12}\cdot\vec v_1)
 \right. \nonumber \\ &&  \left.
+48 (\vec n_{12}\cdot\vec v_1)^3
+\frac{5812}{105} {v_2}^2 (\vec n_{12}\cdot\vec v_2)
+\frac{2056}{21} (\vec n_{12}\cdot\vec v_1)(\vec v_1\cdot\vec v_2)
 \right. \nonumber \\ && \left. 
-\frac{2224}{21} (\vec n_{12}\cdot\vec v_2)(\vec v_1\cdot\vec v_2) 
-\frac{696}{5} (\vec n_{12}\cdot\vec v_1)^2 (\vec n_{12}\cdot\vec v_2)
+\frac{744}{5} (\vec n_{12}\cdot\vec v_1)(\vec n_{12}\cdot\vec v_2)^2
 \right. \nonumber \\ &&  \left.
-\frac{288}{5} (\vec n_{12}\cdot\vec v_2)^3
\right]
 \nonumber \\ &  
+&
\frac{m_1^2 {m_2}^3}{r_{12}^5}
 \left[\frac{6388}{105} V^i-\frac{3172}{21}
   (\vec n_{12}\cdot\vec V)  n_{12}^i \right]
 \nonumber \\ & 
+& 
\frac{m_1^2 {m_2}}{r_{12}^3}
V^i \left[\frac{334}{35}
   {v_1}^4+\frac{654}{35} {v_2}^2 {v_1}^2
-\frac{1336}{35} (\vec v_1\cdot\vec v_2) {v_1}^2
+\frac{292}{35}
   {v_2}^4
 \right. \nonumber \\ && \left. 
-\frac{348}{5} (\vec n_{12}\cdot\vec v_1)^2  V^2 
+ 60 (\vec n_{12}\cdot\vec V)^4
-\frac{1252}{35} {v_2}^2 (\vec v_1\cdot\vec v_2) 
 \right. \nonumber \\ &&  \left. 
+\frac{684}{5} (\vec n_{12}\cdot\vec v_1) (\vec n_{12}\cdot\vec v_2)  V^2 
+\frac{1308}{35} (\vec v_1\cdot\vec v_2)^2  
-66 (\vec n_{12}\cdot\vec v_2)^2  V^2 \right]
 \nonumber 
\end{eqnarray} 
\begin{eqnarray}
&+& 
\frac{m_1^2 {m_2}}{r_{12}^3}
n_{12}^i \left[-\frac{246}{35} (\vec n_{12}\cdot\vec V)  {v_1}^4
-\frac{534}{35}
   {v_2}^2 (\vec n_{12}\cdot\vec v_1) {v_1}^2
 \right.  \nonumber \\ && \left. 
+\frac{90}{7} {v_2}^2 (\vec n_{12}\cdot\vec v_2)
   {v_1}^2
+\frac{1068}{35} (\vec n_{12}\cdot\vec v_1) (\vec v_1\cdot\vec v_2) {v_1}^2
-\frac{984}{35} (\vec n_{12}\cdot\vec v_2) (\vec v_1 \cdot\vec v_2) {v_1}^2
 \right. \nonumber \\&&  \left. 
-\frac{204}{35} {v_2}^4 (\vec n_{12}\cdot\vec v_1)
+60 (\vec n_{12}\cdot\vec v_1)^3 V^2
-56 (\vec n_{12}\cdot\vec V)^5
+\frac{24}{7} {v_2}^4 (\vec n_{12}\cdot\vec v_2)
 \right.  \nonumber \\ &&  \left. 
+\frac{984}{35} {v_2}^2 (\vec n_{12}\cdot\vec v_1) (\vec v_1 \cdot\vec v_2)
-\frac{732}{35} {v_2}^2 (\vec n_{12}\cdot\vec v_2)(\vec v_1\cdot\vec v_2)
-180  (\vec n_{12}\cdot\vec v_1)^2 (\vec n_{12}\cdot\vec v_2) V^2
 \right. \nonumber \\ &&  \left. 
-\frac{1068}{35} (\vec n_{12}\cdot\vec v_1) (\vec v_1\cdot\vec v_2)^2 
+174 (\vec n_{12}\cdot\vec v_1) (\vec n_{12}\cdot\vec v_2)^2 V^2
 \right.  \nonumber \\ &&  \left. 
+\frac{180}{7} (\vec n_{12}\cdot\vec v_2) (\vec v_1 \cdot\vec v_2)^2
-54 V^2(\vec n_{12}\cdot\vec v_2)^3
    \right]
 \nonumber \\ &+& 
\frac{m_1^2{m_2}^2}{{r_{12}^4}}
V^i \left[-\frac{152}{21} {v_1}^2-\frac{1768}{105} {v_2}^2
+\frac{454}{15}  (\vec n_{12}\cdot\vec v_1)^2
 \right. \nonumber \\ && \left.  
+\frac{2864}{105} (\vec v_1\cdot\vec v_2)
-\frac{372}{5} (\vec n_{12}\cdot\vec v_1)(\vec n_{12}\cdot\vec v_2)
+\frac{854}{15} (\vec n_{12}\cdot\vec v_2)^2 \right]
\nonumber \\ &+& 
\frac{m_1^2{m_2}^2}{{r_{12}^4}}
n_{12}^i \left[\frac{1432}{35} (\vec n_{12}\cdot\vec v_1)
   {v_2}^2
-\frac{5752}{105} (\vec n_{12}\cdot\vec v_2) {v_2}^2
-\frac{582}{5} (\vec n_{12}\cdot\vec v_1)^3
 \right. \nonumber \\ && \left. 
+\frac{3568}{105}  (\vec n_{12}\cdot\vec V)v_1^2
-\frac{2864}{35} (\vec n_{12}\cdot\vec v_1)(\vec v_1\cdot\vec v_2)
+\frac{10048}{105} (\vec n_{12}\cdot\vec v_2)(\vec v_1\cdot\vec v_2)
 \right.  \nonumber \\ && \left. 
+\frac{1746}{5} (\vec n_{12}\cdot\vec v_1)^2(\vec n_{12}\cdot\vec v_2)
-\frac{1954}{5} (\vec n_{12}\cdot\vec v_1)(\vec n_{12}\cdot\vec v_2)^2
+158 (\vec n_{12}\cdot\vec v_2)^3 
\right], 
\label{eq:3heom}
\end{eqnarray}
where the acceleration up to the 3 PN order is given in Paper III. 

Eq. (\ref{eq:3heom}) is in perfect agreement with the previous works in
harmonic coordinates \cite{PW02,Nissanke:2004er},  the 
result in the ADMTT coordinate \cite{Jaranowski:1996nv,Konigsdorffer:2003ue}
by a suitable gauge transformation, and also the results from the energy 
balance argument \cite{Iyer:1993xi,Iyer:1995rn,Blanchet:1996vx}.  
We have used the local conservation law of the stress energy tensor of
the matter and the gravitational field and the surface integral approach
to derive our 3.5 PN equations of motion. We have not {\it a priori}
assumed that the star follows a geodesic in any sense. The strong field point
particle limit enables us to realize a point
particle with strong internal gravity without using a Dirac delta
functional. Nissanke et al. \cite{Nissanke:2004er} assumed that a star follows a geodesic
regularized by the Hadamard Parti Finie regularization (or any other
regularization method that gives the same result, such as the dimensional
regularization).  Thereby, the perfect agreement between our present work and
that work \cite{Nissanke:2004er} confirms that a self-gravitating star follows 
the regularized geodesic at least up to the 3.5 PN order inclusively.

\section*{Acknowledgments}
I am grateful to the anonymous referee who carefully read the original 
manuscript and kindly 
gave comments that have substantially improved this paper. 
This paper is a part of the outcome of the Japan Society of the
Promotion of Science (JSPS) Global Center of Excellence (COE) Program (G01): 
Weaving Science Web beyond Particle-Matter Hierarchy at Tohoku
University, Japan.
Extensive use of the algebraic computation software programs   
Mathematica and MathTensor has been made.

\appendix
\section{$\chi$ Part}
\label{sec:chipart}

This section shows the functional expressions 
of $P_{A \chi}^{\tau}$ 
in terms of $m_A, v_A^i, V^i = v_1^i-v_2^i$, and $r_{12}^i$. Here 
we defined $P_{A \chi}^{\mu}$ as 
\begin{eqnarray}
&&
P_{A \chi}^{\mu} \equiv \epsilon^{-4}\int_{B_A}d^3y
\chi^{\mu\tau\alpha\beta}\mbox{}_{,\alpha\beta}.
\label{eq:DefOfPtauChi}
\end{eqnarray}
By the definition of $\chi^{\mu\nu\alpha\beta}\mbox{}_{,\alpha\beta}$,
$$
16 \pi \chi^{\tau\tau\alpha\beta}\mbox{}_{,\alpha\beta}
= (h^{\tau k}h^{\tau l} - h^{\tau\tau}h^{kl})_{,kl},
$$
and thus we can obtain the functional expressions of $P_{A \chi}^{\mu}$
by evaluating surface integrals using the Gauss's law. 
In fact, up to the 3.5 PN order, the definition of 
$P_{A\chi}^{\tau}$ Eq. (\ref{eq:DefOfPtauChi}) gives
\begin{eqnarray}
\lefteqn{
P^{\tau}_{1\chi} = 
\epsilon^4 \mbox{}_4P^{\tau}_{1\chi} +
\epsilon^5 \frac{4 {m_1}^2 {m_2} 
(\vec {n_{12}}\cdot \vec {V})}{3 {r_{12}}^2}
+ \epsilon^6 \mbox{}_6P^{\tau}_{1\chi}} \nonumber \\
&+&
\epsilon^7 \left(
\frac{m_1^3m_2}{r_{12}^3}
\left[
-\frac{80  (\vec n_{12}\cdot \vec v_1) }{9 }
+\frac{92  (\vec {n_{12}}\cdot \vec {v_2}) }{9 }
\right]
-
\frac{16 {m_1}^2{m_2}^2
   (\vec n_{12}\cdot \vec V) }{3 {r_{12}}^3}
\right.
\nonumber \\&&\left.
+\frac{m_1^2m_2}{r_{12}^2}
\left[
\frac{98 {v_1}^2 (\vec n_{12}\cdot \vec v_1) }{45}
+\frac{34 {v_2}^2 (\vec n_{12}\cdot \vec v_1) }{9 }
-\frac{18 (\vec n_{12}\cdot \vec v_1)^3}{5 }
-\frac{242 {v_1}^2
   (\vec n_{12}\cdot \vec v_2) }{45 }
\right.\right.
\nonumber \\&&
\left.\left.
-\frac{46 {v_2}^2 (\vec r_{12}\cdot \vec v_2)
   }{9 }
-\frac{196  (\vec n_{12}\cdot \vec v_1)(\vec v_1\cdot \vec v_2)}{45}
+\frac{80  (\vec n_{12}\cdot \vec v_2)(\vec v_1\cdot \vec v_2)}{9 }
\right.\right. 
\nonumber \\&&
+\left.\left.
\frac{78 (\vec n_{12}\cdot \vec v_1)^2(\vec n_{12}\cdot \vec v_2)
   }{5 }
-{20  (\vec n_{12}\cdot \vec v_1)(\vec n_{12}\cdot \vec
v_2)^2}
+8  (\vec n_{12}\cdot \vec v_2)^3 
\right]
\right). 
\label{Ptchi3hPN}
\end{eqnarray}
The explicit expressions for the 2 PN and 3 PN terms, $\mbox{}_4P^{\tau}_{1\chi}$, 
and $\mbox{}_6P^{\tau}_{1\chi}$, are given in Paper III. The 3.5 PN
$\chi$ part of the star's energy $\mbox{}_7P^{\tau}_{1\chi}$ does 
contribute the 3.5 PN field $\mbox{}_{11}h^{\tau\tau}$ (See
Eq. (\ref{hBttInAp})) 
and affects the 3.5 PN equations of motion.

\section{3.5 PN Landau-Lifshitz Pseudo-tensor}
\label{sec:appendix_tLL}

This section lists the components of the 
Landau-Lifshitz Pseudo-tensor at $O(\epsilon^{11})$ in our ordering  
which are necessary to compute the 3.5 PN equations of motion in 
our formalism.  See the Paper II for    
$\mbox{}_{\le 9}[-16 \pi g t_{LL}^{\mu \nu}]$ 
and the Paper III for 
 $\mbox{}_{10}[-16 \pi g t_{LL}^{\mu \nu}]$.
Note that all the divergence such as $h^{\mu k}\mbox{}_{,k}$ 
in the paper II should be replaced by $- h^{\mu \tau}\mbox{}_{,\tau}$ 
in consistency with the following results. This is simply 
because it is practically much 
easier to evaluate $h^{\mu \tau}\mbox{}_{,\tau}$ 
than $- h^{\mu k}\mbox{}_{,k}$.

The Landau-Lifshitz pseudo-tensor \cite{LL1975}
in terms of $h^{\mu\nu}$ which
satisfies the harmonic condition is 
\begin{eqnarray}
(-16 \pi g)t_{LL}^{\mu\nu} &=&
 g_{\alpha\beta}g^{\gamma\delta}
 h^{\mu\alpha}\mbox{}_{,\gamma}
 h^{\nu\beta}\mbox{}_{,\delta}
+ \frac{1}{2}g^{\mu\nu}g_{\alpha\beta}
 h^{\alpha\gamma}\mbox{} _{,\delta}
 h^{\beta\delta}\mbox{}_{,\gamma}
 -2 g_{\alpha\beta}g^{\gamma {\scriptscriptstyle (} \mu}
 h^{\nu {\scriptscriptstyle )}\alpha}\mbox{}_{,\delta}
 h^{\delta\beta}\mbox{}_{,\gamma}
\nonumber \\
\mbox{} &+& 
\frac{1}{2}\left(g^{\mu\alpha}g^{\nu\beta}
			- \frac{1}{2}g^{\mu\nu}g^{\alpha\beta}
		   \right)
           \left(g_{\gamma\delta}g_{\epsilon\zeta}
			- \frac{1}{2}g_{\gamma\epsilon}g_{\delta\zeta}
		   \right)
			h^{\gamma\epsilon}\mbox{} _{,\alpha}
            h^{\delta\zeta}\mbox{}_{,\beta}.
\end{eqnarray}

We expand 
the deviation field $h^{\mu\nu}$ in a power series 
of $\epsilon$; 
$$
h^{\mu\nu} = \sum_{n=0}\epsilon^{4+n} \mbox{}_{n+4}h^{\mu\nu}.
$$
Paper II showed the lowest order of the field is $\epsilon^4$. 
Using this expansion, we expand $t_{LL}^{\mu\nu}$ 
in $\epsilon$. The point to note is that
1) we raise or lower indexes with the flat metric $\eta^{\mu\nu}$ and 
$\eta_{\mu\nu}$, 2) $\eta^{\tau\tau}= - \epsilon^2$ and
$\eta_{\tau\tau}= - \epsilon^{-2}$, 3) $\mbox{}_5h^{ij}\mbox{}_{,k} = 0$
and $\mbox{}_7h^{\tau\tau}\mbox{}_{,k} = 0$, 4) $\mbox{}_5h^{\tau\mu} =0$.

In the following, repeated alphabetical indexes (i.e. excluding $\tau$) must be summed. 
Indexes in round (square) brackets, $(...)$ (or $[...]$) ,means (anti)-symmetrization on the indexes, and indexes
between vertical bars, $|...|$, are excluded from (anti)-symmetrization.

\noindent 
\begin{eqnarray}
\mbox{}_{11}[-gt_{LL}^{\tau\tau}] &=&
\frac{1}{4}\,\,
\mbox{}_4h^{\tau\tau}\mbox{}_{,\tau} \,\,\,\,{\mbox{}_5h^k\mbox{}_{k,\tau}}
-\frac{3}{4} \,\,{\mbox{}_4h^{\tau\tau}\mbox{}_{,\tau}} \,\,{\mbox{}_7h^{\tau\tau}\mbox{}_{,\tau}}
-\,\,{\mbox{}_7h^{\tau k}\mbox{}_{,\tau}} \,\,{\mbox{}_4h^{\tau\tau}}\mbox{}_{,k}
+\frac{7}{8} \,\,{\mbox{}_7h^{\tau\tau}}
   \,\,{\mbox{}_4h^{\tau\tau,k}} \,\,{\mbox{}_4h^{\tau\tau}}\mbox{}_{,k}
\nonumber \\
&+&\frac{7}{8} 
\mbox{}_5h_{kl} 
\,\,{\mbox{}_4h^{\tau\tau,k}}
   \,\,{\mbox{}_4h^{\tau\tau,l}}
-\frac{7}{4} \,\,{\mbox{}_4h^{\tau\tau}}_{,k} \,\,{\mbox{}_9h^{\tau\tau}}^{,k}
+\,\,{\mbox{}_5h}_{kl,\tau}\,\,{\mbox{}_4h^{\tau k,l}}
+2 \,\,{\mbox{}_4h^{\tau}}_{k,l} \,\,{\mbox{}_7h^{\tau (k,l)}}
\nonumber \\
&+&\frac{1}{4} \,\,{\mbox{}_4h^{\tau\tau,k}} \,\,{\mbox{}_7h}^l\mbox{}_{l,k}.
\end{eqnarray}

\begin{eqnarray}
\mbox{}_{11}[-gt_{LL}^{\tau i}] &=
-&\frac{3}{4} \,\,{\mbox{}_4h^{\tau\tau}\mbox{}_{,\tau}} \,\,{\mbox{}_5h}^i\mbox{}_k \,\,{\mbox{}_4h^{\tau\tau,k}}
+\frac{1}{8}
   \,\,{\mbox{}_7h^{\tau}}^i \,\,{\mbox{}_4h^{\tau\tau}}_{,k} \,\,{\mbox{}_4h^{\tau\tau,k}}
-\frac{1}{4} \,\,{\mbox{}_7h^k\mbox{}_{k,\tau}} \,\,{\mbox{}_4h^{\tau\tau,i}}
\nonumber \\ &
+&\frac{3}{4}
   \,\,{\mbox{}_9h^{\tau\tau}\mbox{}_{,\tau}} \,\,{\mbox{}_4h^{\tau\tau,i}}
-\frac{3}{4} \,\,{\mbox{}_7h^{\tau\tau}\mbox{}_{,\tau}} \,\,{\mbox{}_4h^{\tau\tau}} \,\,{\mbox{}_4h^{\tau\tau,i}}
-\frac{3}{4} \,\,{\mbox{}_4h^{\tau\tau}\mbox{}_{,\tau}} \,\,{\mbox{}_7h^{\tau\tau}}
   \,\,{\mbox{}_4h^{\tau\tau,i}}
\nonumber \\ &
-&\frac{1}{4} \,\,{\mbox{}_7h^{\tau k}}
   \,\,{\mbox{}_4h^{\tau\tau}}\mbox{}_{,k} \,\,{\mbox{}_4h^{\tau\tau,i}}
-\frac{1}{4} \,\,{\mbox{}_5h^k\mbox{}_{k,\tau}} \,\,{\mbox{}_6h^{\tau\tau,i}}
+\frac{3}{4} \,\,{\mbox{}_7h^{\tau\tau}\mbox{}_{,\tau}}
   \,\,{\mbox{}_6h^{\tau\tau,i}}
+\frac{3}{4} \,\,{\mbox{}_4h^{\tau\tau}\mbox{}_{,\tau}} \,\,{\mbox{}_9h^{\tau\tau,i}}
\nonumber \\ &
-& 2\,\,{\mbox{}_7h^{\tau\tau}} \,\,{\mbox{}_4h^{\tau\tau}}\mbox{}_{,k}
   \,\,{\mbox{}_4h^{\tau [k,i]}}
+2 \,\,{\mbox{}_9h^{\tau\tau}}\mbox{}_{,k} \,\,{\mbox{}_4h^{\tau [k,i]}}
-\,\,{\mbox{}_5h}^{ik} \,\,{\mbox{}_4h^{\tau\tau,l}}
   \,\,{\mbox{}_4h^{\tau}}_{l,k}
-\,\,{\mbox{}_7h^{\tau}}\mbox{}_{k,\tau} \,\,{\mbox{}_4h^{\tau}}^{i,k}
\nonumber \\ &
+&\,\,{\mbox{}_5h}_{kl} \,\,{\mbox{}_4h^{\tau\tau,k}}
   \,\,{\mbox{}_4h^{\tau}}^{i,l}
- 2\,\,{\mbox{}_4h^{\tau\tau}} \,\,{\mbox{}_4h^{\tau\tau}}_{,k} \,\,{\mbox{}_7h^{\tau}}^{[k,i]}
+ 2 \,\,{\mbox{}_6h^{\tau\tau}}_{,k} \,\,{\mbox{}_7h^{\tau}}^{[k,i]}
-\,\,{\mbox{}_4h^{\tau}}_{k,\tau} \,\,{\mbox{}_7h^{\tau}}^{i,k}
\nonumber \\ &
+&2 \,\,{\mbox{}_4h^{\tau\tau}}_{,k}
   \,\,{\mbox{}_9h^{\tau}}^{[k,i]}
+\frac{1}{4} \,\,{\mbox{}_5h^k\mbox{}_{k,\tau}} \,\,{\mbox{}_4h}^l\mbox{}_l\mbox{}^{,i}
-\frac{1}{4} \,\,{\mbox{}_7h^{\tau\tau}\mbox{}_{,\tau}} \,\,{\mbox{}_4h}^k\mbox{}_k\mbox{}^{,i}
-2 \,\,{\mbox{}_4h^{\tau}}\mbox{}_{k,l}
   \,\,{\mbox{}_7h}^{k[l,i]}
\nonumber \\ &
-&\frac{1}{2} \,\,{\mbox{}_5h\mbox{}_{kl,\tau}} \,\,{\mbox{}_4h}^{kl,i}
-2 \,\,{\mbox{}_7h^{\tau}}\mbox{}_{k,l}
   \,\,{\mbox{}_4h}^{k[l,i]}
+\,\,{\mbox{}_5h\mbox{}_{kl,\tau}} \,\,{\mbox{}_4h}^{ki,l}
-\frac{1}{4} \,\,{\mbox{}_4h^{\tau\tau}\mbox{}_{,\tau}} \,\,{\mbox{}_7h}^k\mbox{}_k\mbox{}^{,i}.
\end{eqnarray}

\newpage
\begin{eqnarray}
\mbox{}_{11}[-gt_{LL}^{ij}] &=
-&\frac{1}{4} {\delta}^{ij} 
\left(\,\,
{\mbox{}_{11}h^{\tau\tau}}_{,k} \,\,{\mbox{}_4h^{\tau\tau,k}}
+   \,\,{\mbox{}_4h^{\tau\tau}}_{,l} \,\,{\mbox{}_9h^k\mbox{}_k}\mbox{}^{,l}
   \right)
+\frac{1}{2} \,\,{\mbox{}_{11}h^{\tau\tau}}^{,(i}
   \,\,{\mbox{}_4h^{|\tau\tau|}}\mbox{}^{,j)}
+\frac{1}{4} \,\,{\mbox{}_9h^k\mbox{}_k}\mbox{}^{,(i} \,\,{\mbox{}_4h^{|\tau\tau|}}\mbox{}^{,j)} 
\nonumber \\
&+&
\,\,{\mbox{}_9h^{\tau\tau}}\left(
\frac{1}{4}  {\delta}^{ij} \,\,{\mbox{}_4h^{\tau\tau}}_{,k} \,\,{\mbox{}_4h^{\tau\tau}}^{,k}
-\frac{1}{2}  \,\,{\mbox{}_4h^{\tau\tau}}\mbox{}^{,i}
   \,\,{\mbox{}_4h^{\tau\tau}}\mbox{}^{,j}\right)
\nonumber \\
&+& 
{\delta}^{ij}
\,\,{\mbox{}_9h^{\tau\tau}}\mbox{}_{,k}
\left( -\,\,{\mbox{}_4h^{\tau k}\mbox{}_{,\tau}} 
+\frac{1}{2} \,\,{\mbox{}_4h^{\tau\tau}} \,\,{\mbox{}_4h^{\tau\tau,k}}
-\frac{1}{4} \mbox{}_6h^{\tau\tau,k}\mbox{}
-\frac{1}{4} 
   \,\,{\mbox{}_4h}^l\mbox{}_{l}\mbox{}^{,k}
   \right)
\nonumber \\
&+&
2 \,\,{\mbox{}_4h^{\tau (i}\mbox{}_{,\tau}}
   \,\,{\mbox{}_9h^{|\tau\tau|}}\mbox{}^{,j)}
-  \,\,{\mbox{}_4h^{\tau\tau}} \,\,{\mbox{}_4h^{\tau\tau,(i}} \,\,{\mbox{}_9h^{|\tau\tau|,j)}}
+\frac{1}{2} \,\,{\mbox{}_6h^{\tau\tau,(i}}
   \,\,{\mbox{}_9h^{|\tau\tau|,j)}}
+\frac{1}{4} \,\,{\mbox{}_4h}^k\mbox{}_k\mbox{}^{,(i} \,\,{\mbox{}_9h^{|\tau\tau|,j)}} 
\nonumber \\
&-&\frac{3}{4} \,\,{\mbox{}_4h^{\tau\tau}\mbox{}_{,\tau}} \,\,{\mbox{}_9h^{\tau\tau}\mbox{}_{,\tau}} {\delta}^{ij}
\nonumber \\
&-&\,\,{\mbox{}_9h^{\tau k}\mbox{}_{,\tau}} {\delta}^{ij} \,\,{\mbox{}_4h^{\tau\tau}}\mbox{}_{,k}
+ 2 \,\,{\mbox{}_9h^{\tau (j}\mbox{}_{,\tau}} \,\,{\mbox{}_4h^{|\tau\tau|,i)}}
\nonumber \\
&+& 2{\delta}^{ij} 
 \,\,{\mbox{}_4h^{\tau}}_{k,l} \,\,{\mbox{}_9h^{\tau}}^{[k,l]}
\nonumber \\
&-&2 \,\,{\mbox{}_4h^{\tau}}_k\mbox{}^{,(j} \,\,{\mbox{}_9h^{|\tau k|,i)}}
+2 \,\,{\mbox{}_4h^{\tau (j}}\mbox{}_{,|k|} \,\,{\mbox{}_9h^{|\tau k|,i)}}
+2 \,\,{\mbox{}_4h^{\tau}}_k\mbox{}^{,(j}\,\,{\mbox{}_9h^{i) \tau,k}}
-2\,\,{\mbox{}_4h^{\tau}}^{(j}\mbox{}_{,|k|} \,\,{\mbox{}_9h^{i) \tau,k}}
\nonumber \\
&+&
 {\delta}^{ij}
\,\,{\mbox{}_7h^{\tau\tau}}
\left(
 \frac{3}{8}(\,\,{\mbox{}_4h^{\tau\tau}\mbox{}_{,\tau}})^2
+  \,\,{\mbox{}_4h^{\tau k}\mbox{}_{,\tau}} \,\,{\mbox{}_4h^{\tau\tau}}\mbox{}_{,k}
-\frac{3}{4}\,\,{\mbox{}_4h^{\tau\tau}}  \,\,{\mbox{}_4h^{\tau\tau}}\mbox{}_{,k} \,\,{\mbox{}_4h^{\tau\tau,k}}
+\frac{1}{2}  \,\,{\mbox{}_4h^{\tau\tau}}\mbox{}_{,k}\,\,{\mbox{}_6h^{\tau\tau,k}}
\right.
\nonumber \\
&&- \left.
  \,\,{\mbox{}_4h^{\tau}}\mbox{}_{k,l} \,\,{\mbox{}_4h^{\tau [k,l]}}
+\frac{1}{4}  \,\,{\mbox{}_4h^{\tau\tau}}\mbox{}_{,k} \,\,{\mbox{}_4h}^l\mbox{}_l\mbox{}^{,k}
   \right)
\nonumber \\
&+& \,\,{\mbox{}_7h^{\tau\tau}}\left(
- 
2 \,\,{\mbox{}_4h^{\tau\tau,(i}}\,\,{\mbox{}_4h^{j) \tau}\mbox{}_{,\tau}} 
+\frac{3}{2} \,\,{\mbox{}_4h^{\tau\tau}}  \,\,{\mbox{}_4h^{\tau\tau,i}}
   \,\,{\mbox{}_4h^{\tau\tau,j}}
-   \,\,{\mbox{}_6h^{\tau\tau,(i}} \,\,{\mbox{}_4h^{|\tau\tau|,j)}}
\right. 
\nonumber \\
&& \left. 
-\frac{1}{2}
   \,\,{\mbox{}_4h}^k\mbox{}_k\mbox{}^{,(j} \,\,{\mbox{}_4h^{|\tau\tau|,i)}}
+ \,\,{\mbox{}_4h^{\tau}}\mbox{}_k\mbox{}^{,i} \,\,{\mbox{}_4h^{\tau k,j}}
- 2 \,\,{\mbox{}_4h^{\tau}}\mbox{}_k\mbox{}^{,(i} \,\,{\mbox{}_4h^{j) \tau,k}}
+ \,\,{\mbox{}_4h^{\tau i}}\mbox{}_{,k} \,\,{\mbox{}_4h^{\tau j,k}}
   \right)
\nonumber \\
&+& 
{\delta}^{ij} \,\,{\mbox{}_7h^{\tau\tau}\mbox{}_{,\tau}}
\left(
\frac{1}{4}\,\,{\mbox{}_4h^k\mbox{}_{k,\tau}} 
-\frac{3}{4} \,\,{\mbox{}_6h^{\tau\tau}\mbox{}_{,\tau}} 
+\frac{3}{4} \,\,{\mbox{}_4h^{\tau\tau}\mbox{}_{,\tau}} \,\,{\mbox{}_4h^{\tau\tau}}
+\frac{1}{4} \,\,{\mbox{}_4h^{\tau}}_k \,\,{\mbox{}_4h^{\tau\tau}}\mbox{}^{,k} 
   \right)
\nonumber \\
&-&\frac{1}{2} \,\,{\mbox{}_7h^{\tau\tau}\mbox{}_{,\tau}} \,\,{\mbox{}_4h^{\tau\tau,(i}}\,\,{\mbox{}_4h^{j) \tau}} 
\nonumber \\
&+& {\delta}^{ij} \,\,{\mbox{}_7h^{\tau}}_k
\left(
\frac{1}{4} \,\,{\mbox{}_4h^{\tau\tau}\mbox{}_{,\tau}}  \,\,{\mbox{}_4h^{\tau\tau,k}}
-\frac{1}{2} 
   \,\,{\mbox{}_4h^{\tau\tau}}_{,l} \,\,{\mbox{}_4h^{\tau k,l}}
   \right)
\nonumber \\
&
-&\frac{1}{2} \,\,{\mbox{}_4h^{\tau\tau}\mbox{}_{,\tau}} \,\,{\mbox{}_4h^{\tau\tau,(i}} \,\,{\mbox{}_7h^{j) \tau}} 
+ \,\,{\mbox{}_7h^{\tau}}_k \,\,{\mbox{}_4h^{\tau k,(i}}
   \,\,{\mbox{}_4h^{|\tau\tau|,j)}}
\nonumber \\
&+&
{\delta}^{ij} \,\,{\mbox{}_7h^{\tau}\mbox{}_{k,\tau}}
\left(
\,\,{\mbox{}_4h^{\tau\tau}}  \,\,{\mbox{}_4h^{\tau\tau,k}}- 
   \,\,{\mbox{}_6h^{\tau\tau,k}}
   \right)
\nonumber \\
&+& \,\,{\mbox{}_4h^{\tau (i}\mbox{}_{,|\tau|}} \,\,{\mbox{}_7h^{j) \tau}\mbox{}_{,\tau}}
-\,\,{\mbox{}_4h^{\tau\tau}} \,\,{\mbox{}_4h^{\tau\tau,(i}} \,\,{\mbox{}_7h^{j) \tau}\mbox{}_{,\tau}}
+\,\,{\mbox{}_6h^{\tau\tau,(i}}
   \,\,{\mbox{}_7h^{j) \tau }\mbox{}_{,\tau}}
\nonumber \\
&+& 
{\delta}^{ij} \,\,{\mbox{}_7h^{\tau k,l}}
\left(
\,\,{\mbox{}_4h\mbox{}_{kl,\tau}} 
-\frac{1}{2} \,\,{\mbox{}_4h^{\tau}}\mbox{}_k \,\,{\mbox{}_4h^{\tau\tau}}\mbox{}_{,l}
- 2\,\,{\mbox{}_4h^{\tau\tau}} \,\,{\mbox{}_4h^{\tau}}\mbox{}_{[k,l]} 
+2 \,\,{\mbox{}_6h^{\tau}}\mbox{}_{[k,l]}
   \right)
\nonumber \\
&-&
2\,\,{\mbox{}_7h^{\tau k,(i}}\,\,{\mbox{}_4h^{j)}\mbox{}_k\mbox{}_{,\tau}} 
+ \,\,{\mbox{}_4h^{\tau}}\mbox{}_k \,\,{\mbox{}_4h^{\tau\tau,(i}}
   \,\,{\mbox{}_7h^{|\tau k|,j)}}
+2\,\,{\mbox{}_4h^{\tau\tau}} \,\,{\mbox{}_4h^{\tau}}\mbox{}_k\mbox{}^{,(i} \,\,{\mbox{}_7h^{|\tau k|,j)}}
\nonumber \\&
-&
2\,\,{\mbox{}_4h^{\tau\tau}} \,\,{\mbox{}_7h^{\tau k,(i}}\,\,{\mbox{}_4h^{j) \tau}}\mbox{}_{,k} 
-2\,\,{\mbox{}_6h^{\tau}}\mbox{}_k\mbox{}^{,(i} \,\,{\mbox{}_7h^{|\tau k|,j)}}
+2\,\,{\mbox{}_7h^{\tau k,(i }}\,\,{\mbox{}_6h^{j) \tau}}\mbox{}_{,k}
+2\,\,{\mbox{}_6h^{\tau}}\mbox{}_k\mbox{}^{,(i} \,\,{\mbox{}_7h^{j)\tau,k}}
\nonumber \\&
-&
2\,\,{\mbox{}_6h^{\tau,(i}}\mbox{}_{,|k|}
   \,\,{\mbox{}_7h^{j)\tau,k}}
-2\,\,{\mbox{}_4h^{\tau\tau}} \,\,{\mbox{}_4h^{\tau}}\mbox{}_k\mbox{}^{,(i} \,\,{\mbox{}_7h^{j)\tau,k}}
+2\,\,{\mbox{}_4h^{\tau\tau}} \,\,{\mbox{}_4h^{\tau (i}}\mbox{}_{,|k|}
   \,\,{\mbox{}_7h^{j)\tau,k}}
\nonumber 
\end{eqnarray}

\newpage
\begin{eqnarray}
&+&
\frac{1}{8} \,\,{\mbox{}_7h}_{kl} {\delta}^{ij} \,\,{\mbox{}_4h^{\tau\tau,k}} \,\,{\mbox{}_4h^{\tau\tau,l}}
\nonumber \\&+&
\frac{1}{8} \,\,{\mbox{}_7h}^{ij} \,\,{\mbox{}_4h^{\tau\tau}}\mbox{}_{,k} \,\,{\mbox{}_4h^{\tau\tau,k}}
-\frac{1}{2} \,\,{\mbox{}_7h}_k\mbox{}^{(i}
   \,\,{\mbox{}_4h^{|\tau\tau|,j)}} \,\,{\mbox{}_4h^{\tau\tau,k}}
\nonumber \\&+&
{\delta}^{ij} 
\left(
\frac{1}{4} \,\,{\mbox{}_4h^{\tau\tau}} \,\,{\mbox{}_4h^{\tau\tau}}\mbox{}_{,k} \,\,{\mbox{}_7h}^l\mbox{}_l\mbox{}^{,k}
-\frac{1}{4} \,\,{\mbox{}_6h^{\tau\tau}}\mbox{}_{,k}
   \,\,{\mbox{}_7h}^l\mbox{}_l\mbox{}^{,k}
+\frac{1}{4} \,\,{\mbox{}_4h}^k\mbox{}_k\mbox{}_{,l} \,\,{\mbox{}_7h}^m\mbox{}_m\mbox{}^{,l}
-\frac{1}{2} \,\,{\mbox{}_4h}_{kl,m}
   \,\,{\mbox{}_7h}^{kl,m}
+\,\,{\mbox{}_4h}_{kl,m} \,\,{\mbox{}_7h}^{km,l}
   \right)
\nonumber \\&
-&\frac{1}{2} \,\,{\mbox{}_4h^{\tau\tau}} \,\,{\mbox{}_7h}^k\mbox{}_k\mbox{}^{,(i}\,\,{\mbox{}_4h^{|\tau\tau|,j)}} 
+\frac{1}{2}\,\,{\mbox{}_7h}^k\mbox{}_k\mbox{}^{,(i} \,\,{\mbox{}_6h^{|\tau\tau|,j)}}
-\frac{1}{2} \,\,{\mbox{}_4h}^k\mbox{}_k\mbox{}^{,(j} \,\,{\mbox{}_7h}^{|l|}\mbox{}_{|l|}\mbox{}^{,i)}
+ \,\,{\mbox{}_4h}_{kl}\mbox{}^{,(j} \,\,{\mbox{}_7h}^{|kl|,i)}
\nonumber \\&
-&2 \,\,{\mbox{}_7h}^{kl,(i}\,\,{\mbox{}_4h}^{j)}\mbox{}_k\mbox{}_{,l}
-2\,\,{\mbox{}_4h}_{kl}\mbox{}^{,(j} \,\,{\mbox{}_7h}^{i)k,l}
+2\,\,{\mbox{}_7h}^{k(i,|l|} \,\,{\mbox{}_4h}^{j)}\mbox{}_k\mbox{}_{,l}
\nonumber \\ &+&
\,\,{\mbox{}_7h\mbox{}_{kl,\tau}} {\delta}^{ij} \,\,{\mbox{}_4h^{\tau k,l}}
\nonumber \\ &+&
-2\,\,{\mbox{}_4h^{\tau k,(i}}\,\,{\mbox{}_7h^{j)}\mbox{}_{k,\tau}} 
\nonumber \\ &+&
\frac{1}{4} \,\,{\mbox{}_4h^{\tau\tau}\mbox{}_{,\tau}} \,\,{\mbox{}_7h^k\mbox{}_{k,\tau}} {\delta}^{ij}
\nonumber \\ &
-&2 \,\,{\mbox{}_4h^{(j}\mbox{}_{|k,\tau|}} \,\,{\mbox{}_5h^{i)}\mbox{}_{k,\tau}}
+2\,\,{\mbox{}_4h^{\tau\tau,(i}}\,\,{\mbox{}_5h^{j)}\mbox{}_{k,\tau}}\,\,{\mbox{}_4h^{\tau k}} 
- 2\,\,{\mbox{}_6h^{\tau k,(i}}\,\,{\mbox{}_5h^{j)}\mbox{}_{k,\tau}}
\nonumber \\ &+&
{\delta}^{ij}\,\,{\mbox{}_5h^l\mbox{}_{l,\tau}}
\left(
 -\frac{1}{4}\,\,{\mbox{}_4h^k\mbox{}_{k,\tau}}
+\frac{1}{4}\,\,{\mbox{}_6h^{\tau\tau}\mbox{}_{,\tau}}
+\frac{1}{4} \,\,{\mbox{}_4h^{\tau}}\mbox{}_k
   \,\,{\mbox{}_4h^{\tau\tau,k}} 
   \right)
\nonumber \\ &
-&\frac{1}{2} \,\,{\mbox{}_5h^k\mbox{}_{k,\tau}} \,\,{\mbox{}_4h^{\tau\tau,(i}}\,\,{\mbox{}_4h^{j) \tau}} 
\nonumber \\ &+&
{\delta}^{ij} \,\,{\mbox{}_5h\mbox{}_{kl,\tau}}
\left(
\frac{1}{2} \,\,{\mbox{}_4h^{kl}\mbox{}_{,\tau}} 
-\,\,{\mbox{}_4h^{\tau k}} \,\,{\mbox{}_4h^{\tau\tau,l}}
+ \,\,{\mbox{}_6h^{\tau k,l}}
   \right)
\nonumber \\ &
-&2\,\,{\mbox{}_5h^{k(i}\mbox{}_{,|\tau|}} \,\,{\mbox{}_4h^{j)}\mbox{}_{k,\tau}} 
+2\,\,{\mbox{}_4h^{\tau\tau,(i}}\,\,{\mbox{}_5h^{j)}\mbox{}_{k,\tau}}
   \,\,{\mbox{}_4h^{\tau k}} 
-2\,\,{\mbox{}_6h^{\tau k,(i}}\,\,{\mbox{}_5h^{j)}\mbox{}_{k,\tau}} 
\nonumber \\&+&
{\delta}^{ij} \,\,{\mbox{}_5h}_{kl}
\left(
-\frac{1}{4} \,\,{\mbox{}_4h^{\tau\tau}}  \,\,{\mbox{}_4h^{\tau\tau,k}} \,\,{\mbox{}_4h^{\tau\tau,l}}
+\frac{1}{4}
    \,\,{\mbox{}_4h^{\tau\tau,k}} \,\,{\mbox{}_6h^{\tau\tau,l}}
+\frac{1}{2}  \,\,{\mbox{}_4h^{\tau k}}\mbox{}_{,m}
   \,\,{\mbox{}_4h^{\tau l,m}}
\right.
\nonumber \\&& \left.
-\frac{1}{2}  \,\,{\mbox{}_4h^{\tau}}\mbox{}_m\mbox{}^{,k} \,\,{\mbox{}_4h^{\tau m,l}}
+\frac{1}{4}
    \,\,{\mbox{}_4h^{\tau\tau,k}} \,\,{\mbox{}_4h}^m\mbox{}_m\mbox{}^{,l}
-\frac{1}{4}  \,\,{\mbox{}_4h^{\tau\tau}}\mbox{}_{,m}
   \,\,{\mbox{}_4h}^{kl,m}
   \right)
\nonumber \\&+&
\frac{3}{8} \,\,{\mbox{}_5h}^{ij} (\,\,{\mbox{}_4h^{\tau\tau}\mbox{}_{,\tau}})^2
-2 \,\,{\mbox{}_5h}^{(i}\mbox{}_{|k|} \,\,{\mbox{}_4h^{j) \tau}\mbox{}_{,\tau}} 
   \,\,{\mbox{}_4h^{\tau\tau,k}}
+\,\,{\mbox{}_4h^{\tau}\mbox{}_{k,\tau}}
   \,\,{\mbox{}_5h}^{ij} \,\,{\mbox{}_4h^{\tau\tau,k}}
-\frac{1}{4} \,\,{\mbox{}_4h^{\tau\tau}} \,\,{\mbox{}_5h}^{ij} \,\,{\mbox{}_4h^{\tau\tau}}\mbox{}_{,k}
   \,\,{\mbox{}_4h^{\tau\tau,k}}
\nonumber \\&+&
 \,\,{\mbox{}_4h^{\tau\tau}}  \,\,{\mbox{}_4h^{\tau\tau,k}} \,\,{\mbox{}_4h^{\tau\tau,(i}}\,\,{\mbox{}_5h}^{j)}\mbox{}_k
+\frac{1}{4} \,\,{\mbox{}_5h}^{ij} \,\,{\mbox{}_4h^{\tau\tau}}\mbox{}_{,k}
   \,\,{\mbox{}_6h^{\tau\tau,k}}
-\frac{1}{2} \,\,{\mbox{}_4h^{\tau\tau,(i}}\,\,{\mbox{}_5h}^{j)}\mbox{}_k  \,\,{\mbox{}_6h^{\tau\tau,k}}
\nonumber \\&
-&\frac{1}{2} \,\,{\mbox{}_6h^{\tau\tau,(i}}\,\,{\mbox{}_5h}^{j)}\mbox{}_k \,\,{\mbox{}_4h^{\tau\tau}}^{,k}
- \,\,{\mbox{}_5h}^{ij} \,\,{\mbox{}_4h^{\tau}}_{k,l} \,\,{\mbox{}_4h^{\tau}}^{[k,l]}
\nonumber \\&
+&2 \,\,{\mbox{}_4h^{\tau l,(i}}\,\,{\mbox{}_5h}^{j)}\mbox{}_k \,\,{\mbox{}_4h^{\tau}}\mbox{}_l\mbox{}^{,k} 
-\,\,{\mbox{}_5h}_{kl} \,\,{\mbox{}_4h^{\tau k,i}}\,\,{\mbox{}_4h^{\tau l,j}}
- \,\,{\mbox{}_4h^{\tau}}\mbox{}_l\mbox{}^{,k}\,\,{\mbox{}_5h}_k\mbox{}^{(i} \,\,{\mbox{}_4h^{j) \tau,l}}
\nonumber \\&
+&\,\,{\mbox{}_5h}_{kl} \,\,{\mbox{}_4h^{\tau i,k}}\,\,{\mbox{}_4h^{\tau j,l}}
+\frac{1}{4} \,\,{\mbox{}_5h}^{ij} \,\,{\mbox{}_4h^{\tau\tau}}\mbox{}_{,k} \,\,{\mbox{}_4h}^l\mbox{}_l\mbox{}^{,k}
-\frac{1}{2}\,\,{\mbox{}_4h^{\tau\tau,(i}}\,\,{\mbox{}_5h}^{j)}\mbox{}_k \,\, {\mbox{}_4h}^l\mbox{}_l\mbox{}^{,k}
\nonumber \\&
-&\frac{1}{2} \,\,{\mbox{}_4h}^l\mbox{}_l\mbox{}^{,(i}\,\,{\mbox{}_5h}^{j)}\mbox{}_k \,\,{\mbox{}_4h^{\tau\tau,k}} 
+\frac{1}{2} \,\,{\mbox{}_5h}_{kl} \,\,{\mbox{}_4h^{\tau\tau,(j}}
   \,\,{\mbox{}_4h}^{|kl|,i)}.
\end{eqnarray}


\begin{thebibliography}{66}
\expandafter\ifx\csname natexlab\endcsname\relax\def\natexlab#1{#1}\fi
\expandafter\ifx\csname bibnamefont\endcsname\relax
  \def\bibnamefont#1{#1}\fi
\expandafter\ifx\csname bibfnamefont\endcsname\relax
  \def\bibfnamefont#1{#1}\fi
\expandafter\ifx\csname citenamefont\endcsname\relax
  \def\citenamefont#1{#1}\fi
\expandafter\ifx\csname url\endcsname\relax
  \def\url#1{\texttt{#1}}\fi
\expandafter\ifx\csname urlprefix\endcsname\relax\def\urlprefix{URL }\fi
\providecommand{\bibinfo}[2]{#2}
\providecommand{\eprint}[2][]{\url{#2}}

\bibitem[{\citenamefont{Sathyaprakash and Schutz}(2009)}]{Sathyaprakash:2009xs}
\bibinfo{author}{\bibfnamefont{B.~S.} \bibnamefont{Sathyaprakash}}
  \bibnamefont{and} \bibinfo{author}{\bibfnamefont{B.~F.}
  \bibnamefont{Schutz}}, \bibinfo{journal}{Living Rev. Rel.}
  \textbf{\bibinfo{volume}{12}}, \bibinfo{pages}{2} (\bibinfo{year}{2009}).

\bibitem[{\citenamefont{Bildsten and Cutler}(1992)}]{Bildsten:1992my}
\bibinfo{author}{\bibfnamefont{L.}~\bibnamefont{Bildsten}} \bibnamefont{and}
  \bibinfo{author}{\bibfnamefont{C.}~\bibnamefont{Cutler}},
  \bibinfo{journal}{Astrophys. J.} \textbf{\bibinfo{volume}{400}},
  \bibinfo{pages}{175} (\bibinfo{year}{1992}).

\bibitem[{\citenamefont{{Blanchet}}(1997)}]{1997rggr.conf...33B}
\bibinfo{author}{\bibfnamefont{L.}~\bibnamefont{{Blanchet}}}, in
  \emph{\bibinfo{booktitle}{Relativistic Gravitation and Gravitational
  Radiation}}, edited by \bibinfo{editor}{\bibnamefont{{J.-A.~Marck \&
  J.-P.~Lasota}}} (\bibinfo{publisher}{Cambridge University Press, Cambridge},
  \bibinfo{year}{1997}), p.~\bibinfo{pages}{33}.

\bibitem[{\citenamefont{Blanchet}(2006)}]{Blanchet:2006zz}
\bibinfo{author}{\bibfnamefont{L.}~\bibnamefont{Blanchet}},
  \bibinfo{journal}{Living Rev. Rel.} \textbf{\bibinfo{volume}{9}},
  \bibinfo{pages}{4} (\bibinfo{year}{2006}).

\bibitem[{\citenamefont{Ohta et~al.}(1973)\citenamefont{Ohta, Okamura, Kimura,
  and Hiida}}]{Ohta:1973je}
\bibinfo{author}{\bibfnamefont{T.}~\bibnamefont{Ohta}},
  \bibinfo{author}{\bibfnamefont{H.}~\bibnamefont{Okamura}},
  \bibinfo{author}{\bibfnamefont{T.}~\bibnamefont{Kimura}}, \bibnamefont{and}
  \bibinfo{author}{\bibfnamefont{K.}~\bibnamefont{Hiida}},
  \bibinfo{journal}{Prog. Theor. Phys.} \textbf{\bibinfo{volume}{50}},
  \bibinfo{pages}{492} (\bibinfo{year}{1973}).

\bibitem[{\citenamefont{Ohta et~al.}(1974)\citenamefont{Ohta, Okamura, Hiida,
  and Kimura}}]{Ohta:1974pq}
\bibinfo{author}{\bibfnamefont{T.}~\bibnamefont{Ohta}},
  \bibinfo{author}{\bibfnamefont{H.}~\bibnamefont{Okamura}},
  \bibinfo{author}{\bibfnamefont{K.}~\bibnamefont{Hiida}}, \bibnamefont{and}
  \bibinfo{author}{\bibfnamefont{T.}~\bibnamefont{Kimura}},
  \bibinfo{journal}{Prog. Theor. Phys.} \textbf{\bibinfo{volume}{51}},
  \bibinfo{pages}{1220} (\bibinfo{year}{1974}).

\bibitem[{\citenamefont{{Sch{\"a}fer}}(1985)}]{1985AnPhy.161...81S}
\bibinfo{author}{\bibfnamefont{G.}~\bibnamefont{{Sch{\"a}fer}}},
  \bibinfo{journal}{Annals of Physics} \textbf{\bibinfo{volume}{161}},
  \bibinfo{pages}{81} (\bibinfo{year}{1985}).

\bibitem[{\citenamefont{{Damour} and
  {Sch{\"a}fer}}(1985)}]{1985GReGr..17..879D}
\bibinfo{author}{\bibfnamefont{T.}~\bibnamefont{{Damour}}} \bibnamefont{and}
  \bibinfo{author}{\bibfnamefont{G.}~\bibnamefont{{Sch{\"a}fer}}},
  \bibinfo{journal}{General Relativity and Gravitation}
  \textbf{\bibinfo{volume}{17}}, \bibinfo{pages}{879} (\bibinfo{year}{1985}).

\bibitem[{\citenamefont{{Sch{\"a}fer}}(1986)}]{1986GReGr..18..255S}
\bibinfo{author}{\bibfnamefont{G.}~\bibnamefont{{Sch{\"a}fer}}},
  \bibinfo{journal}{General Relativity and Gravitation}
  \textbf{\bibinfo{volume}{18}}, \bibinfo{pages}{255} (\bibinfo{year}{1986}).

\bibitem[{\citenamefont{Jaranowski and Sch{\"a}fer}(1997)}]{Jaranowski:1996nv}
\bibinfo{author}{\bibfnamefont{P.}~\bibnamefont{Jaranowski}} \bibnamefont{and}
  \bibinfo{author}{\bibfnamefont{G.}~\bibnamefont{Sch{\"a}fer}},
  \bibinfo{journal}{Phys. Rev. D} \textbf{\bibinfo{volume}{55}},
  \bibinfo{pages}{4712} (\bibinfo{year}{1997}).

\bibitem[{\citenamefont{{Jaranowski} and {Sch{\"a}fer}}(1998)}]{JS98}
\bibinfo{author}{\bibfnamefont{P.}~\bibnamefont{{Jaranowski}}}
  \bibnamefont{and}
  \bibinfo{author}{\bibfnamefont{G.}~\bibnamefont{{Sch{\"a}fer}}},
  \bibinfo{journal}{\prd} \textbf{\bibinfo{volume}{57}}, \bibinfo{pages}{7274}
  (\bibinfo{year}{1998}).

\bibitem[{\citenamefont{Jaranowski and Sch{\"a}fer}(1998)}]{Jaranowski:1998wy}
\bibinfo{author}{\bibfnamefont{P.}~\bibnamefont{Jaranowski}} \bibnamefont{and}
  \bibinfo{author}{\bibfnamefont{G.}~\bibnamefont{Sch{\"a}fer}},
  \bibinfo{journal}{Phys. Rev. D} \textbf{\bibinfo{volume}{57}},
  \bibinfo{pages}{R5948} (\bibinfo{year}{1998}).

\bibitem[{\citenamefont{Jaranowski and Sch{\"a}fer}(1999)}]{Jaranowski:1999ye}
\bibinfo{author}{\bibfnamefont{P.}~\bibnamefont{Jaranowski}} \bibnamefont{and}
  \bibinfo{author}{\bibfnamefont{G.}~\bibnamefont{Sch{\"a}fer}},
  \bibinfo{journal}{Phys. Rev. D} \textbf{\bibinfo{volume}{60}},
  \bibinfo{pages}{124003} (\bibinfo{year}{1999}).

\bibitem[{\citenamefont{Damour et~al.}(2001{\natexlab{a}})\citenamefont{Damour,
  Jaranowski, and Sch{\"a}fer}}]{Damour:2001bu}
\bibinfo{author}{\bibfnamefont{T.}~\bibnamefont{Damour}},
  \bibinfo{author}{\bibfnamefont{P.}~\bibnamefont{Jaranowski}},
  \bibnamefont{and}
  \bibinfo{author}{\bibfnamefont{G.}~\bibnamefont{Sch{\"a}fer}},
  \bibinfo{journal}{Phys. Lett. B} \textbf{\bibinfo{volume}{513}},
  \bibinfo{pages}{147} (\bibinfo{year}{2001}{\natexlab{a}}).

\bibitem[{\citenamefont{K{\"o}nigsd{\"o}rffer
  et~al.}(2003)\citenamefont{K{\"o}nigsd{\"o}rffer, Faye, and
  Sch{\"a}fer}}]{Konigsdorffer:2003ue}
\bibinfo{author}{\bibfnamefont{C.}~\bibnamefont{K{\"o}nigsd{\"o}rffer}},
  \bibinfo{author}{\bibfnamefont{G.}~\bibnamefont{Faye}}, \bibnamefont{and}
  \bibinfo{author}{\bibfnamefont{G.}~\bibnamefont{Sch{\"a}fer}},
  \bibinfo{journal}{Phys. Rev. D} \textbf{\bibinfo{volume}{68}},
  \bibinfo{pages}{044004} (\bibinfo{year}{2003}).

\bibitem[{\citenamefont{Iyer and Will}(1993)}]{Iyer:1993xi}
\bibinfo{author}{\bibfnamefont{B.~R.} \bibnamefont{Iyer}} \bibnamefont{and}
  \bibinfo{author}{\bibfnamefont{C.~M.} \bibnamefont{Will}},
  \bibinfo{journal}{Phys. Rev. Lett.} \textbf{\bibinfo{volume}{70}},
  \bibinfo{pages}{113} (\bibinfo{year}{1993}).

\bibitem[{\citenamefont{Iyer and Will}(1995)}]{Iyer:1995rn}
\bibinfo{author}{\bibfnamefont{B.~R.} \bibnamefont{Iyer}} \bibnamefont{and}
  \bibinfo{author}{\bibfnamefont{C.~M.} \bibnamefont{Will}},
  \bibinfo{journal}{Phys. Rev. D} \textbf{\bibinfo{volume}{52}},
  \bibinfo{pages}{6882} (\bibinfo{year}{1995}).

\bibitem[{\citenamefont{Blanchet}(1997)}]{Blanchet:1996vx}
\bibinfo{author}{\bibfnamefont{L.}~\bibnamefont{Blanchet}},
  \bibinfo{journal}{Phys. Rev. D} \textbf{\bibinfo{volume}{55}},
  \bibinfo{pages}{714} (\bibinfo{year}{1997}).

\bibitem[{\citenamefont{Gopakumar et~al.}(1997)\citenamefont{Gopakumar, Iyer,
  and Iyer}}]{Gopakumar:1997ng}
\bibinfo{author}{\bibfnamefont{A.}~\bibnamefont{Gopakumar}},
  \bibinfo{author}{\bibfnamefont{B.~R.} \bibnamefont{Iyer}}, \bibnamefont{and}
  \bibinfo{author}{\bibfnamefont{S.}~\bibnamefont{Iyer}},
  \bibinfo{journal}{Phys. Rev. D} \textbf{\bibinfo{volume}{55}},
  \bibinfo{pages}{6030} (\bibinfo{year}{1997}), \bibinfo{note}{erratum-ibid. D
  {\bf 57}:6562 (1998)}.

\bibitem[{\citenamefont{Einstein et~al.}(1938)\citenamefont{Einstein, Infeld,
  and Hoffmann}}]{Einstein:1938yz}
\bibinfo{author}{\bibfnamefont{A.}~\bibnamefont{Einstein}},
  \bibinfo{author}{\bibfnamefont{L.}~\bibnamefont{Infeld}}, \bibnamefont{and}
  \bibinfo{author}{\bibfnamefont{B.}~\bibnamefont{Hoffmann}},
  \bibinfo{journal}{Annals Math.} \textbf{\bibinfo{volume}{39}},
  \bibinfo{pages}{65} (\bibinfo{year}{1938}).

\bibitem[{\citenamefont{Einstein and Infeld}(1940)}]{Einstein:1940mt}
\bibinfo{author}{\bibfnamefont{A.}~\bibnamefont{Einstein}} \bibnamefont{and}
  \bibinfo{author}{\bibfnamefont{L.}~\bibnamefont{Infeld}},
  \bibinfo{journal}{Annals Math.} \textbf{\bibinfo{volume}{41}},
  \bibinfo{pages}{455} (\bibinfo{year}{1940}).

\bibitem[{\citenamefont{{Damour} and {Deruelle}}(1981)}]{1981PhLA...87...81D}
\bibinfo{author}{\bibfnamefont{T.}~\bibnamefont{{Damour}}} \bibnamefont{and}
  \bibinfo{author}{\bibfnamefont{N.}~\bibnamefont{{Deruelle}}},
  \bibinfo{journal}{Physics Letters A} \textbf{\bibinfo{volume}{87}},
  \bibinfo{pages}{81} (\bibinfo{year}{1981}).

\bibitem[{\citenamefont{{Damour}}(1982)}]{Damour82}
\bibinfo{author}{\bibfnamefont{T.}~\bibnamefont{{Damour}}},
  \bibinfo{journal}{C. R. Acad. Sci. Paris II}
  \textbf{\bibinfo{volume}{{294}}}, \bibinfo{pages}{1355}
  (\bibinfo{year}{1982}).

\bibitem[{\citenamefont{{Grishchuk} and
  {Kopeikin}}(1983)}]{1983SvAL....9..230G}
\bibinfo{author}{\bibfnamefont{L.~P.} \bibnamefont{{Grishchuk}}}
  \bibnamefont{and} \bibinfo{author}{\bibfnamefont{S.~M.}
  \bibnamefont{{Kopeikin}}}, \bibinfo{journal}{Soviet Astronomy Letters}
  \textbf{\bibinfo{volume}{9}}, \bibinfo{pages}{230} (\bibinfo{year}{1983}).

\bibitem[{\citenamefont{{Kopeikin}}(1985)}]{1985SvA....29..516K}
\bibinfo{author}{\bibfnamefont{S.~M.} \bibnamefont{{Kopeikin}}},
  \bibinfo{journal}{Soviet Astronomy} \textbf{\bibinfo{volume}{29}},
  \bibinfo{pages}{516} (\bibinfo{year}{1985}).

\bibitem[{\citenamefont{{Thorne} and {Hartle}}(1985)}]{1985PhRvD..31.1815T}
\bibinfo{author}{\bibfnamefont{K.~S.} \bibnamefont{{Thorne}}} \bibnamefont{and}
  \bibinfo{author}{\bibfnamefont{J.~B.} \bibnamefont{{Hartle}}},
  \bibinfo{journal}{\prd} \textbf{\bibinfo{volume}{31}}, \bibinfo{pages}{1815}
  (\bibinfo{year}{1985}).

\bibitem[{\citenamefont{Futamase}(1987)}]{Futamase87}
\bibinfo{author}{\bibfnamefont{T.}~\bibnamefont{Futamase}},
  \bibinfo{journal}{Phys. Rev. D} \textbf{\bibinfo{volume}{36}},
  \bibinfo{pages}{321} (\bibinfo{year}{1987}).

\bibitem[{\citenamefont{Blanchet et~al.}(1998)\citenamefont{Blanchet, Faye, and
  Ponsot}}]{BFP98}
\bibinfo{author}{\bibfnamefont{L.}~\bibnamefont{Blanchet}},
  \bibinfo{author}{\bibfnamefont{G.}~\bibnamefont{Faye}}, \bibnamefont{and}
  \bibinfo{author}{\bibfnamefont{B.}~\bibnamefont{Ponsot}},
  \bibinfo{journal}{Phys. Rev. D} \textbf{\bibinfo{volume}{58}},
  \bibinfo{pages}{124002} (\bibinfo{year}{1998}).

\bibitem[{\citenamefont{Blanchet and
  Faye}(2000{\natexlab{a}})}]{Blanchet:2000nv}
\bibinfo{author}{\bibfnamefont{L.}~\bibnamefont{Blanchet}} \bibnamefont{and}
  \bibinfo{author}{\bibfnamefont{G.}~\bibnamefont{Faye}},
  \bibinfo{journal}{Phys. Lett. A} \textbf{\bibinfo{volume}{271}},
  \bibinfo{pages}{58} (\bibinfo{year}{2000}{\natexlab{a}}).

\bibitem[{\citenamefont{Itoh et~al.}(2000)\citenamefont{Itoh, Futamase, and
  Asada}}]{IFA00}
\bibinfo{author}{\bibfnamefont{Y.}~\bibnamefont{Itoh}},
  \bibinfo{author}{\bibfnamefont{T.}~\bibnamefont{Futamase}}, \bibnamefont{and}
  \bibinfo{author}{\bibfnamefont{H.}~\bibnamefont{Asada}},
  \bibinfo{journal}{Phys. Rev. D} \textbf{\bibinfo{volume}{62}},
  \bibinfo{pages}{064002} (\bibinfo{year}{2000}).

\bibitem[{\citenamefont{{Pati} and {Will}}(2000)}]{PW00}
\bibinfo{author}{\bibfnamefont{M.~E.} \bibnamefont{{Pati}}} \bibnamefont{and}
  \bibinfo{author}{\bibfnamefont{C.~M.} \bibnamefont{{Will}}},
  \bibinfo{journal}{\prd} \textbf{\bibinfo{volume}{62}},
  \bibinfo{pages}{124015} (\bibinfo{year}{2000}).

\bibitem[{\citenamefont{Blanchet and
  Faye}(2001{\natexlab{a}})}]{Blanchet:2000ub}
\bibinfo{author}{\bibfnamefont{L.}~\bibnamefont{Blanchet}} \bibnamefont{and}
  \bibinfo{author}{\bibfnamefont{G.}~\bibnamefont{Faye}},
  \bibinfo{journal}{Phys. Rev. D} \textbf{\bibinfo{volume}{63}},
  \bibinfo{pages}{062005} (\bibinfo{year}{2001}{\natexlab{a}}).

\bibitem[{\citenamefont{Itoh et~al.}(2001)\citenamefont{Itoh, Futamase, and
  Asada}}]{IFA01}
\bibinfo{author}{\bibfnamefont{Y.}~\bibnamefont{Itoh}},
  \bibinfo{author}{\bibfnamefont{T.}~\bibnamefont{Futamase}}, \bibnamefont{and}
  \bibinfo{author}{\bibfnamefont{H.}~\bibnamefont{Asada}},
  \bibinfo{journal}{Phys. Rev. D} \textbf{\bibinfo{volume}{63}},
  \bibinfo{pages}{064038} (\bibinfo{year}{2001}).

\bibitem[{\citenamefont{Pati and Will}(2002)}]{PW02}
\bibinfo{author}{\bibfnamefont{M.~E.} \bibnamefont{Pati}} \bibnamefont{and}
  \bibinfo{author}{\bibfnamefont{C.~M.} \bibnamefont{Will}},
  \bibinfo{journal}{Phys. Rev. D} \textbf{\bibinfo{volume}{65}},
  \bibinfo{pages}{104008} (\bibinfo{year}{2002}).

\bibitem[{\citenamefont{Itoh and Futamase}(2003)}]{Itoh:2003fy}
\bibinfo{author}{\bibfnamefont{Y.}~\bibnamefont{Itoh}} \bibnamefont{and}
  \bibinfo{author}{\bibfnamefont{T.}~\bibnamefont{Futamase}},
  \bibinfo{journal}{Phys. Rev. D} \textbf{\bibinfo{volume}{68}},
  \bibinfo{pages}{121501 (R)} (\bibinfo{year}{2003}).

\bibitem[{\citenamefont{Itoh}(2004{\natexlab{a}})}]{Itoh:2003fz}
\bibinfo{author}{\bibfnamefont{Y.}~\bibnamefont{Itoh}}, \bibinfo{journal}{Phys.
  Rev. D} \textbf{\bibinfo{volume}{69}}, \bibinfo{pages}{064018}
  (\bibinfo{year}{2004}{\natexlab{a}}).

\bibitem[{\citenamefont{Itoh}(2004{\natexlab{b}})}]{Itoh:2004dh}
\bibinfo{author}{\bibfnamefont{Y.}~\bibnamefont{Itoh}},
  \bibinfo{journal}{Class. Quant. Grav.} \textbf{\bibinfo{volume}{21}},
  \bibinfo{pages}{S529} (\bibinfo{year}{2004}{\natexlab{b}}).

\bibitem[{\citenamefont{Nissanke and Blanchet}(2005)}]{Nissanke:2004er}
\bibinfo{author}{\bibfnamefont{S.}~\bibnamefont{Nissanke}} \bibnamefont{and}
  \bibinfo{author}{\bibfnamefont{L.}~\bibnamefont{Blanchet}},
  \bibinfo{journal}{Class. Quant. Grav.} \textbf{\bibinfo{volume}{22}},
  \bibinfo{pages}{1007} (\bibinfo{year}{2005}).

\bibitem[{\citenamefont{{Wang} and {Will}}(2007)}]{2007PhRvD..75f4017W}
\bibinfo{author}{\bibfnamefont{H.}~\bibnamefont{{Wang}}} \bibnamefont{and}
  \bibinfo{author}{\bibfnamefont{C.~M.} \bibnamefont{{Will}}},
  \bibinfo{journal}{\prd} \textbf{\bibinfo{volume}{75}},
  \bibinfo{pages}{064017} (\bibinfo{year}{2007}).

\bibitem[{\citenamefont{{Mitchell} and {Will}}(2007)}]{2007PhRvD..75l4025M}
\bibinfo{author}{\bibfnamefont{T.}~\bibnamefont{{Mitchell}}} \bibnamefont{and}
  \bibinfo{author}{\bibfnamefont{C.~M.} \bibnamefont{{Will}}},
  \bibinfo{journal}{\prd} \textbf{\bibinfo{volume}{75}},
  \bibinfo{pages}{124025} (\bibinfo{year}{2007}).

\bibitem[{\citenamefont{{Damour}}(1983)}]{Damour1983}
\bibinfo{author}{\bibfnamefont{T.}~\bibnamefont{{Damour}}}, in
  \emph{\bibinfo{booktitle}{Gravitational Radiation}}, edited by
  \bibinfo{editor}{\bibfnamefont{N.}~\bibnamefont{{Deruelle}}}
  \bibnamefont{and} \bibinfo{editor}{\bibfnamefont{T.}~\bibnamefont{{Piran}}}
  (\bibinfo{publisher}{NATO Advanced Study Institute, Amsterdam ; New York :
  North-Holland}, \bibinfo{year}{1983}), p.~\bibinfo{pages}{58}.

\bibitem[{\citenamefont{{Schutz}}(1985)}]{Schutz85}
\bibinfo{author}{\bibfnamefont{B.~F.} \bibnamefont{{Schutz}}}, in
  \emph{\bibinfo{booktitle}{Relative Supersymmetry and Cosmology}}, edited by
  \bibinfo{editor}{\bibfnamefont{O.}~\bibnamefont{{Bressan}}},
  \bibinfo{editor}{\bibfnamefont{M.}~\bibnamefont{{Castagnino}}},
  \bibnamefont{and} \bibinfo{editor}{\bibfnamefont{V.}~\bibnamefont{{Hamity}}}
  (\bibinfo{publisher}{World Scientific, Hong Kong}, \bibinfo{year}{1985}),
  p.~\bibinfo{pages}{3}.

\bibitem[{\citenamefont{{Damour}}(1987)}]{Damour1987}
\bibinfo{author}{\bibfnamefont{T.}~\bibnamefont{{Damour}}}, in
  \emph{\bibinfo{booktitle}{Three hundred years of gravitation}}, edited by
  \bibinfo{editor}{\bibfnamefont{S.~W.} \bibnamefont{{Hawking}}}
  \bibnamefont{and} \bibinfo{editor}{\bibfnamefont{W.}~\bibnamefont{{Israel}}}
  (\bibinfo{publisher}{Cambridge University Press, Cambridge},
  \bibinfo{year}{1987}), p. \bibinfo{pages}{128}.

\bibitem[{\citenamefont{{Asada} and {Futamase}}(1997)}]{AsadaFutamase97}
\bibinfo{author}{\bibfnamefont{H.}~\bibnamefont{{Asada}}} \bibnamefont{and}
  \bibinfo{author}{\bibfnamefont{T.}~\bibnamefont{{Futamase}}},
  \bibinfo{journal}{Progress of Theoretical Physics Supplement}
  \textbf{\bibinfo{volume}{128}}, \bibinfo{pages}{123} (\bibinfo{year}{1997}).

\bibitem[{\citenamefont{Futamase and Itoh}(2007)}]{Futamase:2007zz}
\bibinfo{author}{\bibfnamefont{T.}~\bibnamefont{Futamase}} \bibnamefont{and}
  \bibinfo{author}{\bibfnamefont{Y.}~\bibnamefont{Itoh}},
  \bibinfo{journal}{Living Rev. Rel.} \textbf{\bibinfo{volume}{10}},
  \bibinfo{pages}{2} (\bibinfo{year}{2007}).

\bibitem[{\citenamefont{Blanchet and Faye}(2000{\natexlab{b}})}]{BF00}
\bibinfo{author}{\bibfnamefont{L.}~\bibnamefont{Blanchet}} \bibnamefont{and}
  \bibinfo{author}{\bibfnamefont{G.}~\bibnamefont{Faye}}, \bibinfo{journal}{J.
  Math. Phys.} \textbf{\bibinfo{volume}{41}}, \bibinfo{pages}{7675}
  (\bibinfo{year}{2000}{\natexlab{b}}).

\bibitem[{\citenamefont{Blanchet and Faye}(2001{\natexlab{b}})}]{BF01}
\bibinfo{author}{\bibfnamefont{L.}~\bibnamefont{Blanchet}} \bibnamefont{and}
  \bibinfo{author}{\bibfnamefont{G.}~\bibnamefont{Faye}}, \bibinfo{journal}{J.
  Math. Phys.} \textbf{\bibinfo{volume}{42}}, \bibinfo{pages}{4391}
  (\bibinfo{year}{2001}{\natexlab{b}}).

\bibitem[{\citenamefont{de~Andrade et~al.}(2001)\citenamefont{de~Andrade,
  Blanchet, and Faye}}]{deAndrade:2000gf}
\bibinfo{author}{\bibfnamefont{V.~C.} \bibnamefont{de~Andrade}},
  \bibinfo{author}{\bibfnamefont{L.}~\bibnamefont{Blanchet}}, \bibnamefont{and}
  \bibinfo{author}{\bibfnamefont{G.}~\bibnamefont{Faye}},
  \bibinfo{journal}{Class. Quant. Grav.} \textbf{\bibinfo{volume}{18}},
  \bibinfo{pages}{753} (\bibinfo{year}{2001}).

\bibitem[{\citenamefont{Damour et~al.}(2001{\natexlab{b}})\citenamefont{Damour,
  Jaranowski, and Sch{\"a}fer}}]{Damour:2000ni}
\bibinfo{author}{\bibfnamefont{T.}~\bibnamefont{Damour}},
  \bibinfo{author}{\bibfnamefont{P.}~\bibnamefont{Jaranowski}},
  \bibnamefont{and}
  \bibinfo{author}{\bibfnamefont{G.}~\bibnamefont{Sch{\"a}fer}},
  \bibinfo{journal}{Phys. Rev. D} \textbf{\bibinfo{volume}{63}},
  \bibinfo{pages}{044021} (\bibinfo{year}{2001}{\natexlab{b}}).

\bibitem[{\citenamefont{Blanchet et~al.}(2004)\citenamefont{Blanchet, Damour,
  and Esposito-Farese}}]{Blanchet:2003gy}
\bibinfo{author}{\bibfnamefont{L.}~\bibnamefont{Blanchet}},
  \bibinfo{author}{\bibfnamefont{T.}~\bibnamefont{Damour}}, \bibnamefont{and}
  \bibinfo{author}{\bibfnamefont{G.}~\bibnamefont{Esposito-Farese}},
  \bibinfo{journal}{Phys. Rev. D} \textbf{\bibinfo{volume}{69}},
  \bibinfo{pages}{124007} (\bibinfo{year}{2004}).

\bibitem[{\citenamefont{Fukumoto et~al.}(2006)\citenamefont{Fukumoto, Futamase,
  and Itoh}}]{Fukumoto:2006gv}
\bibinfo{author}{\bibfnamefont{T.}~\bibnamefont{Fukumoto}},
  \bibinfo{author}{\bibfnamefont{T.}~\bibnamefont{Futamase}}, \bibnamefont{and}
  \bibinfo{author}{\bibfnamefont{Y.}~\bibnamefont{Itoh}},
  \bibinfo{journal}{Prog. Theor. Phys.} \textbf{\bibinfo{volume}{116}},
  \bibinfo{pages}{423} (\bibinfo{year}{2006}).

\bibitem[{\citenamefont{{Schutz}}(1980)}]{Schutz80}
\bibinfo{author}{\bibfnamefont{B.~F.} \bibnamefont{{Schutz}}},
  \bibinfo{journal}{\prd} \textbf{\bibinfo{volume}{22}}, \bibinfo{pages}{249}
  (\bibinfo{year}{1980}).

\bibitem[{\citenamefont{{Futamase} and {Schutz}}(1983)}]{FS83}
\bibinfo{author}{\bibfnamefont{T.}~\bibnamefont{{Futamase}}} \bibnamefont{and}
  \bibinfo{author}{\bibfnamefont{B.~F.} \bibnamefont{{Schutz}}},
  \bibinfo{journal}{\prd} \textbf{\bibinfo{volume}{28}}, \bibinfo{pages}{2363}
  (\bibinfo{year}{1983}).

\bibitem[{\citenamefont{{Futamase}}(1983)}]{Futamase83}
\bibinfo{author}{\bibfnamefont{T.}~\bibnamefont{{Futamase}}},
  \bibinfo{journal}{\prd} \textbf{\bibinfo{volume}{28}}, \bibinfo{pages}{2373}
  (\bibinfo{year}{1983}).

\bibitem[{\citenamefont{Futamase and Schutz}(1985)}]{FS85}
\bibinfo{author}{\bibfnamefont{T.}~\bibnamefont{Futamase}} \bibnamefont{and}
  \bibinfo{author}{\bibfnamefont{B.~F.} \bibnamefont{Schutz}},
  \bibinfo{journal}{Phys. Rev. D} \textbf{\bibinfo{volume}{32}},
  \bibinfo{pages}{2557} (\bibinfo{year}{1985}).

\bibitem[{\citenamefont{{Futamase}}(1985)}]{Futamase85}
\bibinfo{author}{\bibfnamefont{T.}~\bibnamefont{{Futamase}}},
  \bibinfo{journal}{\prd} \textbf{\bibinfo{volume}{32}}, \bibinfo{pages}{2566}
  (\bibinfo{year}{1985}).

\bibitem[{\citenamefont{{Anderson} and {Decanio}}(1975)}]{AndersonDecanio1975}
\bibinfo{author}{\bibfnamefont{J.~L.} \bibnamefont{{Anderson}}}
  \bibnamefont{and} \bibinfo{author}{\bibfnamefont{T.~C.}
  \bibnamefont{{Decanio}}}, \bibinfo{journal}{General Relativity and
  Gravitation} \textbf{\bibinfo{volume}{6}}, \bibinfo{pages}{197}
  (\bibinfo{year}{1975}).

\bibitem[{\citenamefont{Fock}(1959)}]{Fock1959}
\bibinfo{author}{\bibfnamefont{V.~A.} \bibnamefont{Fock}},
  \emph{\bibinfo{title}{{Theory of Space, Time and Gravitation}}}
  (\bibinfo{publisher}{Pergamon Press, London}, \bibinfo{year}{1959}).

\bibitem[{\citenamefont{{Ehlers} et~al.}(1976)\citenamefont{{Ehlers},
  {Rosenblum}, {Goldberg}, and {Havas}}}]{Ehlers1976}
\bibinfo{author}{\bibfnamefont{J.}~\bibnamefont{{Ehlers}}},
  \bibinfo{author}{\bibfnamefont{A.}~\bibnamefont{{Rosenblum}}},
  \bibinfo{author}{\bibfnamefont{J.~N.} \bibnamefont{{Goldberg}}},
  \bibnamefont{and} \bibinfo{author}{\bibfnamefont{P.}~\bibnamefont{{Havas}}},
  \bibinfo{journal}{Astrophysical Journal} \textbf{\bibinfo{volume}{208}},
  \bibinfo{pages}{L77} (\bibinfo{year}{1976}).

\bibitem[{\citenamefont{{Dixon}}(1979)}]{Dixon1979}
\bibinfo{author}{\bibfnamefont{W.~G.} \bibnamefont{{Dixon}}}, in
  \emph{\bibinfo{booktitle}{Isolated Gravitating Systems in General
  Relativity}}, edited by \bibinfo{editor}{\bibnamefont{{J.~Ehlers}}}
  (\bibinfo{publisher}{North-Holland Pub. Co., Amsterdam},
  \bibinfo{year}{1979}), p. \bibinfo{pages}{156}.

\bibitem[{\citenamefont{{Ashby} and {Bertotti}}(1986)}]{AB86}
\bibinfo{author}{\bibfnamefont{N.}~\bibnamefont{{Ashby}}} \bibnamefont{and}
  \bibinfo{author}{\bibfnamefont{B.}~\bibnamefont{{Bertotti}}},
  \bibinfo{journal}{\prd} \textbf{\bibinfo{volume}{34}}, \bibinfo{pages}{2246}
  (\bibinfo{year}{1986}).

\bibitem[{\citenamefont{{Will} and {Wiseman}}(1996)}]{WW96}
\bibinfo{author}{\bibfnamefont{C.~M.} \bibnamefont{{Will}}} \bibnamefont{and}
  \bibinfo{author}{\bibfnamefont{A.~G.} \bibnamefont{{Wiseman}}},
  \bibinfo{journal}{\prd} \textbf{\bibinfo{volume}{54}}, \bibinfo{pages}{4813}
  (\bibinfo{year}{1996}).

\bibitem[{\citenamefont{{Blanchet} and {Damour}}(1986)}]{BlanchetDamour86}
\bibinfo{author}{\bibfnamefont{L.}~\bibnamefont{{Blanchet}}} \bibnamefont{and}
  \bibinfo{author}{\bibfnamefont{T.}~\bibnamefont{{Damour}}},
  \bibinfo{journal}{Royal Society of London Philosophical Transactions Series
  A} \textbf{\bibinfo{volume}{320}}, \bibinfo{pages}{379}
  (\bibinfo{year}{1986}).

\bibitem[{\citenamefont{{Poujade} and {Blanchet}}(2002)}]{PoujadeBlanchet2002}
\bibinfo{author}{\bibfnamefont{O.}~\bibnamefont{{Poujade}}} \bibnamefont{and}
  \bibinfo{author}{\bibfnamefont{L.}~\bibnamefont{{Blanchet}}},
  \bibinfo{journal}{\prd} \textbf{\bibinfo{volume}{65}},
  \bibinfo{pages}{124020} (\bibinfo{year}{2002}).

\bibitem[{\citenamefont{{Landau} and {Lifshitz}}(1975)}]{LL1975}
\bibinfo{author}{\bibfnamefont{L.~D.} \bibnamefont{{Landau}}} \bibnamefont{and}
  \bibinfo{author}{\bibfnamefont{E.~M.} \bibnamefont{{Lifshitz}}},
  \emph{\bibinfo{title}{{The classical theory of fields 4th rev. engl. ed}}}
  (\bibinfo{publisher}{Pergamon Press, Oxford}, \bibinfo{year}{1975}).

\bibitem[{\citenamefont{{Blanchet} and {Damour}}(1988)}]{BD88}
\bibinfo{author}{\bibfnamefont{L.}~\bibnamefont{{Blanchet}}} \bibnamefont{and}
  \bibinfo{author}{\bibfnamefont{T.}~\bibnamefont{{Damour}}},
  \bibinfo{journal}{\prd} \textbf{\bibinfo{volume}{37}}, \bibinfo{pages}{1410}
  (\bibinfo{year}{1988}).

\end{thebibliography}
\end{document}